\documentclass[reprint, nofootinbib, aps, prd, superscriptaddress]{revtex4-2}

\usepackage{amsmath}

\DeclareMathOperator{\arcosh}{arcosh}
\usepackage{amssymb}
\usepackage{hyperref}
\usepackage{booktabs}
\usepackage{mathtools}
\usepackage[dvipsnames]{xcolor}

\newcommand{\p}{\partial}

\renewcommand{\AA}{\mathcal{A}}

\newcommand{\JJ}{\mathcal{J}}

\newcommand{\xperp}{\mathbf x}
\newcommand{\yperp}{\mathbf y}
\newcommand{\pperp}{\mathbf p}
\newcommand{\qperp}{\mathbf q}
\newcommand{\kperp}{\mathbf k}
\newcommand{\uperp}{\mathbf u}
\newcommand{\vperp}{\mathbf v}

\newcommand{\sperp}{\mathbf s}
\newcommand{\zeroperp}{\mathbf 0}
\newcommand{\nn}{\nonumber}

\newcommand{\xp}{x^+}
\newcommand{\xm}{x^-}

\newcommand{\tr}{\mathrm{Tr}}

\newcommand{\elrf}{\epsilon_\mathrm{LRF}}
\newcommand{\ii}{i} 
\newcommand{\ind}{\hspace{0.4cm}}

\begin{document}

\title{Energy-momentum tensor of the dilute (3+1)D Glasma}

\date{\today}

\author{Andreas Ipp}
\email{ipp@hep.itp.tuwien.ac.at}
\affiliation{Institute for Theoretical Physics, TU Wien, Wiedner Hauptstraße  8-10/136, A-1040 Vienna, Austria}
\author{Markus Leuthner}
\email{mleuthner@hep.itp.tuwien.ac.at}
\affiliation{Institute for Theoretical Physics, TU Wien, Wiedner Hauptstraße  8-10/136, A-1040 Vienna, Austria}
\author{David I.~M\"uller}
\email{dmueller@hep.itp.tuwien.ac.at}
\affiliation{Institute for Theoretical Physics, TU Wien, Wiedner Hauptstraße  8-10/136, A-1040 Vienna, Austria}
\author{S\"oren Schlichting}
\email{sschlichting@physik.uni-bielefeld.de}
\affiliation{Fakult\"at f\"ur Physik, Universit\"at Bielefeld, D-33615 Bielefeld, Germany}
\author{Kayran Schmidt}
\email{kschmidt@hep.itp.tuwien.ac.at}
\affiliation{Institute for Theoretical Physics, TU Wien, Wiedner Hauptstraße  8-10/136, A-1040 Vienna, Austria}
\author{Pragya Singh}
\email{pragya.phy.singh@jyu.fi}
\affiliation{Department of Physics, University of Jyv\"askyl\"a, FI-40014 Jyv\"askyl\"a, Finland}
\affiliation{Helsinki Institute of Physics, University of Helsinki, FI-00014 Helsinki, Finland}

\begin{abstract}
We present a succinct formulation of the energy-momentum tensor of the Glasma characterizing the initial color fields in relativistic heavy-ion collisions in the Color Glass Condensate effective theory. We derive concise expressions for the (3+1)D dynamical evolution of symmetric nuclear collisions in the weak field approximation employing a generalized McLerran-Venugopalan model with non-trivial longitudinal correlations.
Utilizing Monte Carlo integration, we calculate in unprecedented detail non-trivial rapidity profiles of early-time observables at RHIC and LHC energies, including transverse energy densities and eccentricities.
For our setup with broken boost invariance, we carefully discuss the placement of the origin of the Milne frame and interpret the components of the energy-momentum tensor.
We find longitudinal flow that deviates from standard Bjorken flow in the (3+1)D case and provide a geometric interpretation of this effect.
Furthermore, we observe a universal shape in the flanks of the rapidity profiles regardless of collision energy and predict that limiting fragmentation should also hold at LHC energies.
\end{abstract}

\maketitle

\section{Introduction}
Relativistic heavy-ion collisions (HIC) are studied experimentally at the Large Hadron Collider (LHC) and Relativistic Heavy Ion Collider (RHIC) as a method to probe quantum chromodynamics (QCD) at extreme energies~\cite{ Heinz:2013th, Gale:2013da, Busza:2018rrf}, where quarks and gluons are liberated from color confinement and form the Quark Gluon Plasma (QGP).
The intricate spacetime evolution of the system, coupled with changing degrees of freedom, presents a difficult challenge to first principles calculations and phenomenological model building.
The current state-of-the-art understanding of the creation and evolution of the medium relies on sophisticated multi-stage modeling, typically split into a pre-equilibrium stage, a hydrodynamic stage, and a hadron gas stage~\cite{Heinz:2013wva}.
Bayesian analysis in the context of multi-state examination reveals correlations between initial conditions and medium properties~\cite{Bernhard:2019bmu, Nijs:2020roc, JETSCAPE:2020mzn}, but it is not clear to what extent the properties of the initial state of a heavy-ion collision can be deduced from experimentally measured final state observables.

Despite these efforts, our current knowledge of the initial conditions in three spatial dimensions for high-energy heavy-ion collisions remains notably limited. The Color Glass Condensate (CGC)~\cite{Gelis:2010nm, Gelis:2012ri} emerges as a promising theoretical framework for providing first principles insight into the initial state, wherein the initial interaction of nuclei is characterized in terms of large and small Bjorken-$x$ partons, leading to a dense state of gluonic matter known as the Glasma~\cite{Lappi:2006fp}. Due to high gluon occupancy, the leading-order dynamics are governed by the Classical Yang-Mills (CYM) equations. The Glasma is thus characterized by strong classical color fields.

In contrast to phenomenological models, such as e.g. MC Glauber-type models \cite{Miller:2007ri}, which aim to directly model the energy density of the collision medium, the CGC/Glasma approach provides a description for nuclei before the collision, as well as the collision itself, and the subsequent pre-equilibrium evolution of the Glasma. 
Since the properties of the Glasma are fully determined by the CYM dynamics and the cold nuclear matter properties of atomic nuclei before the collisions, the modeling aspect of the pre-equilibrium stage is relegated to the distribution of color charge of high-energy nuclei. The IP-Glasma~\cite{Schenke:2012wb, Schenke:2012hg, Gale:2012rq} is one such nuclear model with notable phenomenological success.

Most CGC-based initial state models neglect the longitudinal dynamics of the medium due to the assumption of boost invariance at high energies, effectively rendering the Glasma a (2+1)D system. While boost invariance provides a good approximation for central collisions at mid-rapidity,
recent experimental measurements carried out at the LHC, such as the observation of flow decorrelations~\cite{CMS:2015xmx, ATLAS:2017rij} present a challenge to the assumption of boost invariance in the initial state. Consequently, it becomes essential to delve into the longitudinal dynamics of high-energy collisions. Over the years, various implementations of (3+1)D initial state models have been developed, employing either static~\cite{Schenke:2016ksl, McDonald:2018wql, McDonald:2023qwc} or dynamical sources~\cite{Gelfand:2016yho, Ipp:2017lho, Ipp:2018hai, Ipp:2020igo, Schlichting:2020wrv, Matsuda:2023gle, Matsuda:2024moa} in the CGC framework, or phenomenologically by considering constituent quarks, color flux tubes or strings of varying lengths~\cite{Werner:2010aa, Monnai:2015sca, Bozek:2015tca, Denicol:2015nhu, Shen:2022oyg}.
However, these approaches are either phenomenological in nature or severely constrained by the computational resources required. 

In this paper, we present the extension of the first analytical calculation of energy deposition in high-energy heavy-ion collisions obtained by solving the (3+1)D CYM equations within the weak field approximation~\cite{Ipp:2021lwz}. The results exhibit excellent agreement with fully non-perturbative lattice simulations that share a similar origin of rapidity dependence~\cite{Ipp:2021lwz, Ipp:2022lid}. We now employ the derived analytic expressions for the perturbative gauge fields on the future light cone to formulate the energy-momentum tensor of the Glasma. A key focus of this study lies in presenting a concise expression for the field strength tensor of the Glasma, wherein we rigorously reduce the number of integrals from six to three, optimizing efficiency for Monte Carlo integration. Subsequently, we employ these expressions in conjunction with a three-dimensional generalization of the MV model at two different collision energies corresponding to RHIC and LHC ranges to examine the rapidity profiles of the energy-momentum tensor, local rest frame (LRF) energy density and flow. 

In Sec.~\ref{sec:weak-field} we discuss the weak field approximation in the CGC and show how to obtain the Glasma field strength tensor and Glasma energy-momentum tensor up to leading order in the sources.
The resulting integrals need to be solved numerically, for which we explain our Monte Carlo (MC) integration method.
Next, we introduce a shifted Milne frame for evaluating the energy-momentum tensor of the dilute Glasma and derived observables in the future light cone.
In Sec.~\ref{sec:nuclear_model} we explain the details of our three-dimensional nuclear model, which includes a correlation length parameter that allows us to set the scale of longitudinal fluctuations in our nuclei.
In Sec.~\ref{sec:results} we discuss our numerical results for several observables of the dilute Glasma.
We summarize our findings and give a brief outlook in Sec.~\ref{sec:conclusion}.

\section{Weak field approximation}%
\label{sec:weak-field}

The CGC is an effective field theory for high energy QCD applicable to relativistic heavy-ion collisions \cite{Gelis:2010nm, Gelis:2012ri}. The hard degrees of freedom within the two nuclei are modeled as classical currents of color charge moving at the speed of light. In contrast, the soft gluon fields are sourced by the nuclei and determined by the Classical Yang-Mills equations. While the classical approximation is warranted due to the high gluon occupation numbers and weak coupling at high energies, it should also be noted that it agrees with the quantum treatment to leading order in perturbation theory.

\begin{figure}
    \centering
    \includegraphics{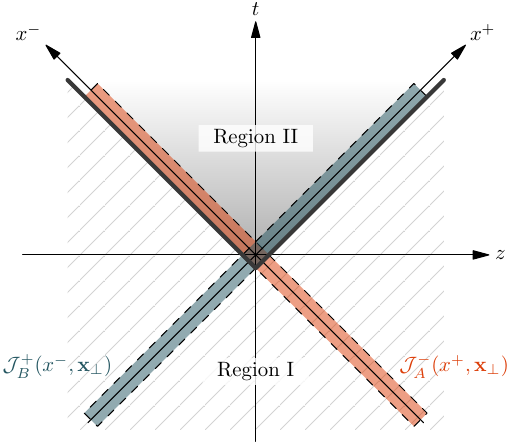}
    \caption{Spacetime diagram of a relativistic heavy-ion collision showing time $t$ and beam axis $z$. The tracks of nuclei $A$ ($B$) are marked in orange (blue). The diamond-shaped region where they overlap is the interaction region, and together with its causal future, it is called \textbf{Region II}. The complement region, which represents the system before the collision, is called \textbf{Region I}.}
    \label{fig:minkowski}
\end{figure}

We use the CGC to describe the initial states and collision of two relativistic nuclei. The geometry of our setup is shown in Fig.~\ref{fig:minkowski}. First, consider a single nucleus $A$ moving in negative $z$-direction through the (pre-collision) \textbf{Region I} at the speed of light. The corresponding color current in light cone coordinates $x^\pm = (x^0 \pm x^3)/\sqrt{2} = (t\pm z)/\sqrt{2}$ is
\begin{align}
    \label{eq:J_A}
    \mathcal J^\mu_A(x^+,\xperp) = \delta^\mu_- \rho_A(x^+,\xperp).
\end{align}
The boldface symbol $\xperp$ refers to the transverse coordinates $\xperp = (x^1, x^2) =  (x,y)$ and $\rho_A(x^+,\xperp)$ is the color charge density, which takes values in the Lie algebra $\mathfrak{su}(N_c)$. Due to Lorentz contraction, the charge density is localized in a thin sheet around $x^+\!=0$. We allow for a finite longitudinal thickness of the nucleus as opposed to the ultrarelativistic (boost-invariant) case, where the nucleus is considered to be infinitesimally thin. Within our treatment, the nucleus is still considered a static source, i.e., the current does not depend on $x^-$ and propagates at the fixed speed of light. Corrections to this behavior provide another source of sub-eikonal corrections~\cite{Lam:2000nz, Ozonder:2012vw, Ozonder:2013moa, Altinoluk:2014oxa, Altinoluk:2015gia}.

In Eq.~\eqref{eq:J_A} and in the following, we always assume covariant gauge, which for the gauge field $\mathcal A^\mu_A$ reads $\p_\mu \mathcal A^\mu_A = 0$.
The Yang-Mills (YM) equations are
\begin{align}
    \label{eq:cym_A}
    \mathcal{D}_\mu \mathcal{F}^{\mu\nu}_A = \p_\mu \mathcal{F}^{\mu\nu}_A - \ii g[\mathcal A_{\mu}^A, \mathcal F^{\mu\nu}_A] =\mathcal J^\nu_A,
\end{align}
with the field strength tensor
\begin{align} \label{eq:field_strength}
    \mathcal{F}_A^{\mu\nu} = \p^\mu \AA^\nu - \p^\nu \AA^\mu -\ii g[\AA^\mu,\AA^\nu].
\end{align}
Using the boundary conditions
\begin{align}
    \mathcal A_A^+ = \mathcal A_A^i= 0,
\end{align}
where $i=x,y$ labels the two transverse components, the YM equations are solved by
\begin{align}
    \label{eq:A_poisson}
    \mathcal{A}^-_A(x^+,\xperp) = -(\nabla_\perp^2)^{-1} \rho_A(x^+,\xperp)=: \phi_A(x^+,\xperp),
\end{align}
with $\nabla_\perp$ acting in the transverse plane. This gauge field and the current satisfy the continuity equation
\begin{align}
\mathcal D_\mu \mathcal J^\mu_A (x^+,\xperp) = 0.
\end{align}
Analogously, introducing a nucleus $B$ moving in the opposite direction with current
\begin{align}
    \mathcal J^\mu_B(x^-,\xperp)=\delta^\mu_+\rho_B(x^-,\xperp),
\end{align}
yields
\begin{align}
    \label{eq:B_poisson}
    \mathcal{A}^+_B(x^-,\xperp) = -(\nabla_\perp^2)^{-1} \rho_B(x^-,\xperp)=: \phi_B(x^-,\xperp).
\end{align}

\subsection{Dilute limit of the Glasma field}

 We now review our treatment of the collision of nuclei $A$ and $B$ in terms of the dilute approximation (see \cite{Ipp:2021lwz, Ipp:2022lid} for details). In the interaction region, where the two nuclei overlap, and its causal future, denoted together in Fig.~\ref{fig:minkowski} as \textbf{Region II}, the solutions in Eqs.~\eqref{eq:A_poisson} and \eqref{eq:B_poisson} are no longer valid and neither is a superposition. Instead, the solution to the full collision problem has to be obtained from the Classical Yang-Mills equations
\begin{align}
    \label{eq:cym_full}
    D_\mu F^{\mu\nu} =J^\nu.
\end{align}
Here, non-calligraphic symbols refer to the interacting solutions of the field equations.
The corresponding gauge field and current
\begin{align}
    \label{eq:A_general}
    A^\mu(x^+,x^-,\xperp) 
    &= \mathcal A^\mu_A(x^+,\xperp) + \mathcal A^\mu_B(x^-,\xperp) + a^\mu(x),\\
    \label{eq:J_general}
    J^\mu(x^+,x^-,\xperp) &= \mathcal J^\mu_A(x^+,\xperp) + \mathcal J^\mu_B(x^-,\xperp)+ j^\mu(x),
\end{align}
are defined in terms of the non-interacting solutions before the collision plus additional correction terms $a^\mu$ and $j^\mu$, which are zero prior to the collision and depend on $x= (x^+,x^-, \xperp)$. Therefore, in the asymptotic past
\begin{align}
    \lim_{t\rightarrow -\infty} a^\mu(x) = 0,\\
    \label{eq:init_j}
    \lim_{t\rightarrow -\infty} j^\mu(x) = 0.
\end{align}

To make further progress, we consider an expansion of Eqs.~\eqref{eq:A_general} and \eqref{eq:J_general} in powers of $\rho_A$ and $\rho_B$. Since we work in the covariant gauge, $\p_\mu a^\mu = 0$. Note that the single nuclei gauge fields $\mathcal A^\mu_{A/B}$ are linear functionals of $\rho_{A/B}$ as expressed in Eqs.~\eqref{eq:A_poisson} and \eqref{eq:B_poisson}, i.e.~they are of order $\rho_{A/B}$. Similarly, the currents $\mathcal{J}^\mu_{A/B}$ are also of order $\rho_{A/B}$. If $\rho_B = 0$, then $A^\mu = \mathcal{A}_A^\mu$ and $J^\mu = \mathcal{J}^\mu_A$ solve Eq.~\eqref{eq:cym_full} to all orders in $\rho_A$. The analog is true for $A^\mu = \mathcal A_B^\mu$ and $J^\mu = \mathcal J_B^\mu$ if $\rho_A = 0$. Therefore, $a^\mu$ and $j^\mu$ capture corrections of order $\rho_A^n \rho_B^m$ with $n,m\geq 1$ in the full collision problem. Expanding Eq.~\eqref{eq:cym_full} in orders of $\rho_A$ and $\rho_B$ the contributions of order $\rho_A\rho_B$ are
\begin{align}
    \label{eq:cym_expansion}
    \p_\mu \left( \p^\mu a^\nu - \ii g[\AA_A^\mu,\AA^\nu_B]+ ig[\AA^\nu_A, \AA_B^\mu]\right)& \nn \\
    - \ii g[\AA^A_\mu,\p^\mu \AA^\nu_B-\p^\nu\AA^\mu_B]& \nn \\
    - ig[\AA^B_\mu, \p^\mu\AA^\nu_A-\p^\nu\AA^\mu_A] &= j^\nu,
\end{align}
where we take $a^\nu$ to be of order $\rho_A\rho_B$ with all higher order terms removed. 
Dropping terms of higher order in $\rho_{A/B}$, we implicitly assume weak sources. Therefore, we refer to this approximation as the dilute limit or weak field approximation. Considering only the contributions of order $\rho_A\rho_B$ to $D_\mu J^\mu = 0$ yields
\begin{align}
    \label{eq:current_conservation_expansion}
    \p_\mu j^\mu = ig[\AA^A_\mu, \JJ^\mu_B] + ig[\AA^B_\mu,\JJ^\mu_A].
\end{align}
 As the nuclei are in recoilless, lightlike motion, their trajectory does not change. Therefore, $j^\mu$ only represents a rotation in color space of the nucleus' currents and is limited to $j^-$ and $j^+$ components, which are localized in the same region as $\rho_A$ and $\rho_B$, respectively. Equation~\eqref{eq:current_conservation_expansion} then splits into two contributions
\begin{align}
    \partial_- j^- = ig[\AA_\mu^B,\JJ^\mu_A],\\
    \partial_+ j^+ = ig[\AA^A_\mu, \JJ^\mu_B].
\end{align}
Given the initial conditions in Eq.~\eqref{eq:init_j} they are solved by
\begin{align}
    j^-(x) = ig\intop_{-\infty}^{x^-}dv^-\, \left[\phi_B(v^-,\xperp),\rho_A(x^+,\xperp)\right],\\
    j^+(x) = ig\intop_{-\infty}^{x^+}dv^+\,\left[\phi_A(v^+,\xperp),\rho_B(x^-,\xperp)\right],
\end{align}
where $\phi_{A/B}$ correspond to the gauge fields of nuclei $A/B$ in Eqs.~\eqref{eq:A_poisson} and~\eqref{eq:B_poisson}.
These expressions can be used in Eq.~\eqref{eq:cym_expansion} to solve for $a^\mu(x)$, which we identify as the Glasma field. We recite the solutions from \cite{Ipp:2021lwz},
\begin{widetext}
\begin{align}\label{eq:ap_final} 
a^+(x) & =  \frac{g}{2} f_{abc} t^c  \intop_{\pperp,\qperp}  \intop_0^{\infty}   dv^+   \intop_0^{\infty} dv^- \tilde \phi^a_A( x^+\! - v^+\!, \pperp) \tilde \phi^b_B(x^-\! - v^-\! , \qperp) \frac{
\big( - (\pperp + \qperp)^2 + 2 \qperp^2 \big) v^+
}{|\pperp + \qperp | \tau'}
J_1( |\pperp + \qperp | \tau') e^{-\ii(\pperp+ \qperp)\cdot \xperp},\\
\label{eq:am_final} 
a^-(x) &= \frac{g}{2} f_{abc} t^c \intop_{\pperp ,\mathbf \qperp}\intop_0^{\infty} dv^+ \intop_0^{\infty} dv^- \tilde \phi^a_A( x^+\! - v^+\!,\pperp) \tilde \phi^b_B(x^-\! - v^-\!, \qperp)\frac{\big(+(\pperp + \qperp)^2 - 2 \pperp^2\big) v^-}{ |\pperp+\qperp|\tau'}
J_1( |\pperp + \qperp | \tau') e^{-\ii(\pperp+ \qperp)\cdot \xperp},
\end{align}
\begin{align}
\label{eq:ai_final} 
    a^i(x) &= i\frac{g}{2} f_{abc} t^c \intop_{\pperp ,\qperp}\intop_0^{\infty} dv^+ \intop_0^{\infty} dv^- \tilde \phi^a_A( x^+\! - v^+\!,\pperp) \tilde \phi^b_B(x^-\! - v^-\!, \qperp)(p^i- q^i)~J_0( |\pperp + \qperp | \tau') e^{-\ii(\pperp+ \qperp)\cdot \xperp}.
\end{align}
\end{widetext}
Here, $f_{abc}$ are the structure constants of the $\mathfrak{su}(N_c)$ Lie algebra and $t^c$ are its generators. The symbol $|\pperp|$ denotes the modulus of the transverse vector $\pperp$, and $\pperp \cdot \xperp = p_i x_i = p^i x^i$.
Quantities with a tilde are understood as Fourier transforms in the transverse plane, e.g.
\begin{align}
    \phi^a_A(x^+\!-v^+\!,\xperp) = \intop_{\pperp} \tilde{\phi}^a_A(x^+\!-v^+\!,\pperp)e^{-\ii\pperp\cdot\xperp},
\end{align}
where $\int_{\pperp} = \int\frac{d^2\pperp}{(2\pi)^2}$. Also, $\tau' = \sqrt{2v^+v^-}$, and $J_m$ are the Bessel functions of the first kind. We note that these solutions are only correct inside the future light cone, i.e.~outside the tracks of the nuclei, as there are additional corrections inside the tracks. However, these are co-moving with the color fields of the single nuclei, such that these corrections do not contribute to the color field of the Glasma in the future light cone and thus can be ignored.

\subsection{Field strength tensor}
The next step is to compute the leading order contribution to the field strength tensor $f^{\mu\nu}$ from the leading order gauge fields $a^\mu$. In general, the non-Abelian field strength tensor (cf.~Eq.~\eqref{eq:field_strength}) contains an Abelian part $\partial^\mu a^\nu - \partial^\nu a^\mu$ and a commutator term $[a^\mu, a^\nu]$. In the dilute limit, the perturbative Glasma fields $a^\mu$ are of order $\rho_A\rho_B$, and the commutator term only contributes to higher-order corrections.
Thus, to leading order, the Glasma field strength tensor can be defined as\footnote{This definition of $f^{\mu\nu}$ differs from the one given in \cite{Ipp:2021lwz}, where contributions from the background fields were included. However, these contributions vanish outside the overlap region of the nuclei.}
\begin{align}
\label{eq:dilute_f}
    f^{\mu\nu} \equiv \p^\mu a^\nu - \p^\nu a^\mu.
\end{align}
After a lengthy derivation, which can be found in Appendix~\ref{sec:derivation}, we arrive at a remarkably simple result. The components of the leading order  field strength tensor are given by
\begin{align}
f^{+-}(x) &= -\frac{g}{2\pi} \intop_{ \eta', \vperp} V(x,  \eta', \vperp), \label{eq:f+-}\\
f^{+i}(x) &=  \frac{g}{2\pi}  \intop_{ \eta' ,\vperp}\big(V^{ij}(x,  \eta', \vperp)-\delta^{ij}V(x,  \eta', \vperp) \big) w^j \frac{e^{+ \eta'}}{\sqrt 2} , \label{eq:f+i}\\
f^{-i}(x) &=  \frac{g}{2\pi}  \intop_{ \eta',\vperp} \big(V^{ij}(x,  \eta', \vperp)+\delta^{ij}V(x,  \eta', \vperp) \big)   w^j \frac{e^{- \eta'}}{\sqrt 2} , \label{eq:f-i}\\
f^{ij}(x) &= -\frac{g}{2\pi} \intop_{ \eta', \vperp} V^{ij}(x,  \eta', \vperp),\label{eq:fij}
\end{align}
where the rapidity-like integration variable \mbox{$\eta' \in (-\infty,\infty)$}, and the $\vperp$-integral spans the entire transverse plane, $\int_\vperp = \int d^2 \vperp$.
The symbols $V$ and $V^{ij}$ are defined as
\begin{align}
\label{eq:def_V}
    V(x,  \eta', \vperp) &= f_{abc} \, t^c \, \beta^{i,a}_A(x,  \eta', \vperp)  \,  \beta^{i,b}_B (x,  \eta', \vperp) ,\\
\label{eq:def_Vij}
    V^{ij}(x,  \eta', \vperp)  &= f_{abc} \, t^c \big( \beta^{i,a}_A(x,  \eta', \vperp) \,  \beta^{j,b}_B(x,  \eta', \vperp) \nonumber \\
    &\hspace{1.5cm} -\beta^{j,a}_A(x,  \eta', \vperp) \, \beta^{i,b}_B(x,  \eta', \vperp) \big),
\end{align}
where $\beta^i$ are the gradients of the color potentials $\phi$,
\begin{align}
    \beta^{i,a}_{A/B}(x,  \eta', \vperp) 
    &= \partial^i_{(x)}\phi^{a}_{A/B}(x^\pm\!- \frac{|\vperp|}{\sqrt{2}}e^{\pm  \eta'}, \xperp-\vperp).
\end{align}
In covariant gauge, these gradients may be identified with the only non-trivial components of the field strength tensors of the single nuclei,
\begin{align}
    \mathcal F^{i-}_A&= \partial^i \mathcal A^-_A=\beta^i_A,\\
    \mathcal F^{i+}_B &= \partial^i \mathcal A^+_B = \beta^i_B.
\end{align}
The vector components $w^j$ are given by
\begin{align}
    w^j = -\p^j_{(v)} |\vperp| =  \frac{v^j}{|\vperp|}.
\end{align}
These concise expressions for the Glasma field strength tensor are our main analytical result and we stress that the boost-invariant limit $\beta^{i,a}_{A/B}(x^\pm, \xperp) = \delta(x^\pm)\alpha^{i,a}_{A/B}(\xperp)$ of our expressions is consistent with previous boost-invariant results \cite{Guerrero-Rodriguez:2021ask, Demirci:2023ejg}.  

\begin{figure}
    \centering
    \includegraphics{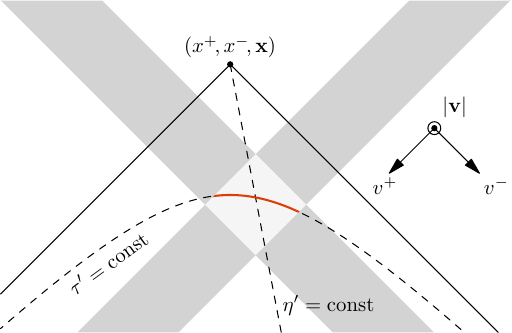}
    \caption{Geometric illustration of the integration paths for the field strength tensor $f^{\mu\nu}$.
    The integration domain is the past light cone of the spacetime point $(\xp, \xm, \xperp)$, parametrized by the Milne coordinates $(\tau'\!=\!|\vperp|, \eta', \vperp)$.
    For a slice at fixed transverse displacement $\tau' = \mathrm{const}$, we highlight the conic section inside the collision region 
    that contributes to the integral (orange solid line).}
    \label{fig:backwards-milne}
\end{figure}

Our solutions for the field strength tensor have an intuitive geometric interpretation illustrated in Fig.~\ref{fig:backwards-milne}.
Given the spacetime point $x$, at which we evaluate $f^{\mu\nu}$, the integration is restricted to the causal past of $x$ along lightlike paths emitted from the collision region.
To see this, we note that the modulus of $\vperp$ in Eqs.~\eqref{eq:f+-} - \eqref{eq:fij} can be interpreted as a time coordinate $ \tau' = |\vperp|$ analogous to Milne proper time $\tau$. Together with the rapidity-like coordinate $ \eta'$, we find that the pair $( \tau',  \eta')$ can be viewed as Milne coordinates for a slice with fixed $|\vperp|$ of the past light cone attached to the spacetime point $x$. 
We emphasize that the displacement four-vector $v^\mu$ 
\begin{align}
    v^\pm = \frac{ \tau'}{\sqrt{2}} e^{\pm  \eta'}, \qquad \vperp = \begin{pmatrix}v^x\\v^y\end{pmatrix}
\end{align}
measures the spacetime distance between the source of emission and the point $x$. Since this vector is lightlike, i.e.~$v_\mu v^\mu = 0$, it is apparent that the three-dimensional integrals in Eqs.~\eqref{eq:f+-} - \eqref{eq:fij} only sum over lightlike paths ending at $x$.
Furthermore, we note that the integrands are non-zero only in the four-dimensional spacetime volume where the two colliding nuclei overlap. Intuitively, this region corresponds to points in spacetime from which gluons are emitted due to interactions among the colliding color fields. These gluons then propagate along lightlike paths into the future light cone until they arrive at $x$, forming the Glasma field strength at that particular point.

Compared to the gauge fields in Eqs.~\eqref{eq:ap_final}-\eqref{eq:ai_final}, the components of the field strength tensor $f^{\mu\nu}$ are much easier to evaluate numerically.
Firstly, the gauge fields are six-dimensional integrals, whereas the field strength tensor only involves three-dimensional integrals. This simplification is possible by exploiting the closure relation of the Bessel functions $J_m$, see Eq.~\eqref{eq:bessel_closure} in the Appendix for details.
Secondly, having eliminated the Bessel functions, the integrands of $f^{\mu\nu}$ are now free of oscillating terms. This way, the numerical evaluation and convergence of Eqs.~\eqref{eq:f+-} - \eqref{eq:fij} is greatly simplified.

\subsection{Monte Carlo integration} \label{sec:mc_sampling}

Despite the relatively simple structure of the expressions, the integrals for the components of the field strength tensor have to be solved numerically in practical computations of the (3+1)D Glasma. In this work, we focus on a Monte Carlo integration approach. We consider the prototypical integral
\begin{align} \label{eq:proto_integral}
    I(x^+,x^-,\xperp) &= \intop_{-\infty}^{+\infty} d  \eta'\! \intop\!d^2 \vperp  \, h(u^+, u^-, \uperp), \\
     u^\pm &= x^\pm - v^\pm = x^\pm - \frac{|\vperp|}{\sqrt{2}}e^{\pm \eta'}, \label{eq:u_lc} \\
     \uperp &= \xperp-\vperp,
\end{align}
where we assume that the function $h(u^+,u^-,\uperp)$ has compact support along the two light cone directions, \mbox{$u^{\pm}_\mathrm{min} < u^\pm < u^\pm_\mathrm{max}$} with longitudinal widths \mbox{$u^\pm_\mathrm{max} - u^\pm_\mathrm{min} > 0$}. Imposing compact support introduces a slight error because the color fields $\phi_{A/B}(x^\pm, \xperp)$, which enter the function $h$, generally only exhibit exponential decay in both longitudinal and transverse directions.  
The constants $u^\pm_\mathrm{max/min}$ can nonetheless be chosen large enough such that our numerical results are barely affected by such a cut-off. Similarly, we restrict the transverse integration domain over $\uperp$ to a rectangular region $u^i_\mathrm{min} < u^i < u^i_\mathrm{max}$ with $u^i_\mathrm{max} > u^i_\mathrm{min}$, where the color fields overlap. Furthermore, we evaluate the integral $I$ only in the region where Eqs.~\eqref{eq:f+-}-\eqref{eq:fij} are valid, that is for $x^\pm > u^\pm_\mathrm{max}$. With these strong assumptions, the integrals are guaranteed to converge.%
\footnote{The conditions for convergence can be weakened if we assume that the color potentials $\phi_{A/B}(u^\pm, \uperp)$ are not compact along the light cone directions, but merely decay to zero for $|u^\pm| \rightarrow \infty$. In this case, the overlap of the two nuclei shown in Fig.~\ref{fig:backwards-milne} is unbounded and the evaluation point $x$ is always inside the overlap region. The convergence of $f^{+-}$ and $f^{ij}$ is guaranteed if the color fields do not exhibit any singularities and decay faster than $|u^\pm|^{-s}$ with $s>1$ as $|u^\pm| \rightarrow \infty$. On the other hand, the components $f^{\pm i}$ in Eqs.~\eqref{eq:f+i} and \eqref{eq:f-i} impose a stronger decay due to the presence of the exponential factors $e^{\pm \eta'}$. Rewriting these integrals using light cone coordinates $v^\pm$ yields $e^{\pm \eta'} \propto \sqrt{v^\pm / v^\mp}$ (see Appendix \ref{sec:derivation}). To ensure convergence for $v^\mp \rightarrow \infty$, the color potentials $\phi_{A/B}$ have to asymptotically decrease as $|u^\pm|^{-s}$ with $s>3/2$. Since nuclear models 
impose at least exponential decay, these conditions are always fulfilled. Nonetheless, we stress that
the derivation of Eqs.~\eqref{eq:f+-}-\eqref{eq:fij} in principle assumes compact support. Evaluating the integrals inside the overlap yields an incomplete result for the Glasma field strength.}

\begin{figure}
    \centering
    \includegraphics{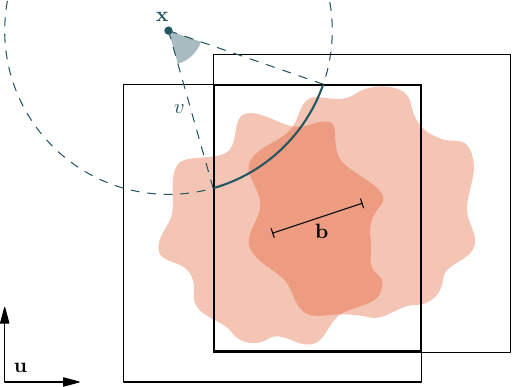}
    \caption{Illustration of two partially overlapping nuclei (orange blobs) separated by the impact parameter $\mathbf{b}$ in the transverse plane spanned by $\uperp$. To evaluate the field strength tensor at $\xperp$, we restrict the angular integration with radius $v$ to the arc (thick blue) within the rectangular region where the two nuclei overlap.}
    \label{fig:mc-integration}
\end{figure}

Our integration strategy is based on rewriting the integral in Eq.~\eqref{eq:proto_integral} using a coordinate transformation and making use of the restricted integration domain. Since Monte Carlo integration relies on random sampling, sub-optimal sampling strategies can lead to unnecessary evaluations of the integrand in regions where it is zero.
We can increase efficiency by determining strict bounds on the coordinate ranges. 
First, we use polar coordinates $(v, \theta)$ for the transverse integration over $\vperp$,
\begin{align} 
    &I(x^+,x^-,\xperp) = \nonumber \\ &\intop_{-\infty}^{+\infty} d  \eta'\! \intop^\infty_0\!d v \, v \intop^{2\pi}_{0} d \theta \, h(x^+\!\!-\! \frac{v}{\sqrt{2}}e^{+ \eta'}, x^-\!\!-\!\frac{v}{\sqrt{2}}e^{- \eta'}, \xperp - \vperp),
\end{align}
where $\vperp = v \, \mathbf{e}(\theta)$ with the unit vector in the $v$-direction \mbox{$\mathbf{e}(\theta) = (\cos \theta, \sin \theta)$}. Secondly, the aforementioned restrictions imposed on the integration domain allow us to put bounds on the integrals over $ \eta'$, $v$, and $\theta$. Considering  only the restrictions on $u^+$, we find that $ \eta'$ is restricted to $(\eta^+_\mathrm{min}(v), \eta^+_\mathrm{max}(v))$ with
\begin{align}
    \eta^+_\mathrm{min/max}(v) &= +\ln  \frac{\sqrt{2} (x^+ - u^+_\mathrm{max/min})}{v} 
\end{align}
for a given value of $v > 0$.
Analogously, we find a restriction from $u^-$ as \mbox{$ \eta' \in (\eta^-_\mathrm{min}(v), \eta^-_\mathrm{max}(v))$} with
\begin{align}
    \eta^-_\mathrm{min/max}(v) &= - \ln \frac{\sqrt{2} (x^- - u^-_\mathrm{min/max})}{v} .
\end{align}
Combining both inequalities for $ \eta'$, the integration domain must be restricted to $ \eta' \in (\eta_\mathrm{min}, \eta_\mathrm{max})$ with \mbox{$\eta_\mathrm{min} = \max(\eta^+_\mathrm{min}, \eta^-_\mathrm{min})$} and \mbox{$\eta_\mathrm{max} = \min(\eta^+_\mathrm{max}, \eta^-_\mathrm{max})$}.
We also deduce bounds on the integration variable $v$. From Eq.~\eqref{eq:u_lc} it follows that
\begin{align}
    v = \sqrt{2 (x^+ - u^+)(x^- - u^-)}.
\end{align}
Saturating the restrictions on $u^\pm$, we find that
\mbox{$v \in (v_\mathrm{min}, v_\mathrm{max})$} with
\begin{align}
    v_\mathrm{min/max} = \sqrt{2 (x^+ - u^+_\mathrm{max/min}) (x^- - u^-_\mathrm{max/min})}.
\end{align}
The angle $\theta$ is also restricted. Given valid choices of $ \eta'$ and $v$, the angular integration corresponds to a circle of radius $v$ with center $\xperp$ in the transverse plane, intersecting the rectangular integration domain only at certain points. Only those segments of the circle that lie inside the restricted integration domain contribute to the integral, as illustrated in  Fig.~\ref{fig:mc-integration}, where the intersection of the circle with the integration domain is a single arc. Determining these arcs then allows for more efficient sampling of $\theta$.

We perform random sampling as follows: we first choose a radius $v$ with probability density $p(v) \propto v$ inside the bounds $(v_\mathrm{min}, v_\mathrm{max})$. Given $v$, we evaluate the bounds on $ \eta'$ and sample uniformly from the interval given by $\eta_\mathrm{min}(v)$ and $\eta_\mathrm{max}(v)$. Finally, we determine where the circle defined by $v$ and $\xperp$ intersects the rectangular domain and obtain the segments from which we sample $\theta$ uniformly. Drawing $M$ samples for all three coordinates, we approximate the integral by
\begin{align} 
    &I(x^+,x^-,\xperp) \approx \nonumber \\
    & \frac{C}{M}\sum^M_{i=1} h(x^+\!\!-\! \frac{v^{(i)}}{\sqrt{2}}e^{+ \eta'^{(i)}}, x^-\!\!-\!\frac{v^{(i)}}{\sqrt{2}}e^{- \eta'^{(i)}}, \xperp - \vperp^{(i)}),
\end{align}
where $\vperp^{(i)} = v^{(i)} (\cos \theta^{(i)}, \sin \theta^{(i)})$ and $C$ is a Jacobian factor, which is proportional to the integration volume.

\subsection{Energy-momentum tensor}%
\label{sec:emt_theory}

From the Glasma field strength tensor given in Eqs.~\eqref{eq:f+-}--\eqref{eq:fij}, it is straightforward to obtain the Glasma energy-momentum tensor
\begin{align}
\label{eq:EMT}
    T^{\mu\nu} = 2\,\tr [f^{\mu\rho} f_{\rho}{}^{\nu} + \frac{1}{4} g^{\mu\nu} f^{\rho\sigma} f_{\rho\sigma}]
\end{align}
in the future light cone. In addition to the energy-momentum tensor, we compute the local rest frame (LRF) energy density $\elrf$ and fluid velocity $u^\mu$ by solving the Landau condition
\begin{align} \label{eq:landau}
    T^\mu{}_\nu u^\nu = \elrf u^\mu
\end{align}
with $u^\mu$ the only timelike eigenvector of $T^{\mu}{}_\nu$ and $\elrf$ its eigenvalue.

\begin{figure}
    \centering
    \includegraphics{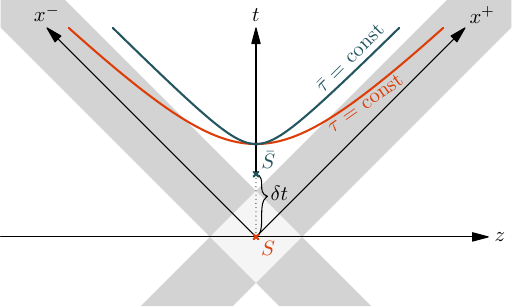}
    \caption{Constant proper time hyperbolas for two different choices of the Milne frame origin separated by the distance $\delta t$. For the origin $S$ (orange), the hyperbola enters the trajectories of the nuclei (gray). This is never the case for the Milne frame centered at $\bar S$ (blue).}
    \label{fig:shift}
\end{figure}

We parametrize the future light cone in Milne coordinates
\begin{align}
\tau&=\sqrt{t^2-z^2},\\
\eta_s&= \frac{1}{2} \ln\frac{t+z}{t-z},
\end{align}
where $\tau$ and $\eta_s$ are proper time and spacetime rapidity, respectively.
Contrary to the boost-invariant setup, it is a priori not clear where the origin of the Milne frame should be placed. As depicted in Fig.~\ref{fig:shift}, putting the origin at the center of the interaction region ($S$) will lead to some of the $\tau=\mathrm{const}$ hyperbolas cutting into the tracks of the nuclei. In order to avoid this, we shift the origin of our Milne frame forward in time by some offset $\delta t$ that is slightly larger than the nuclear radius in the longitudinal direction. None of the $\bar\tau=\mathrm{const}$ hyperbolas with respect to the new origin $\bar S$ cross the trajectories of the nuclei. We evaluate all observables in terms of this shifted Milne frame, but we omit the bars in $\bar\tau$ and $\bar\eta_s$ from now on to reduce clutter.

\section{Nuclear model}\label{sec:nuclear_model}

We study a three-dimensional generalization of the McLerran-Venugopalan (MV) model \cite{McLerran:1993ni, McLerran:1993ka}.
We assume a Gaussian probability functional for the color charges of our nuclei,
propagating either in $x^+$ or in $x^-$ direction,
which is completely fixed by the 1- and 2-point functions. The 1-point function in our model vanishes,
\begin{align}
    \langle \rho^a(x^\pm, \xperp) \rangle = 0.
\end{align}
The 2-point function is given by
\begin{align}
    \langle \rho^a(x^\pm,\xperp) \rho^b(y^\pm,\yperp)\rangle &= g^2\mu^2\delta^{ab}\sqrt{T(x^\pm, \xperp)}\sqrt{T(y^\pm, \yperp)}\nn\\
    &\quad\times U_\xi(x^\pm-y^\pm)\delta^{(2)}(\xperp-\yperp)
    \label{eq:correlator_factorized}
\end{align}
with the MV scale $\mu$, which, in the non-perturbative case, is related to the saturation momentum $Q_s \approx g^2 \mu$.
Here,
\begin{align}
    T(x^\pm, \xperp) = \frac{c}{1+\exp(\frac{\sqrt{2(\gamma_\mathrm{beam} \, x^\pm)^2+\xperp^2}-R}{d})}
\end{align}
is a boosted Woods-Saxon (WS) profile with WS radius $R$ and skin depth parameter $d$. The Lorentz contraction factor $\gamma_\mathrm{beam}$ is determined by the beam velocity and the normalization constant $c$ ensures
\begin{align}
\label{eq:T_norm}
\intop_{-\infty}^\infty dx^\pm\, T(x^\pm, \zeroperp) = 1.
\end{align}
The function $T$ determines the overall shape of the nuclei.
On the other hand, fluctuations within the nuclei are controlled by the function
\begin{align}
    \label{eq:Uxi}
    U_\xi(x^\pm-y^\pm) = \frac{1}{\sqrt{2\pi \xi^2}} e^\frac{(x^\pm-y^\pm)^2}{8R_l^2}e^{-\frac{(x^\pm - y^\pm)^2}{2\xi^2}},
\end{align}
where $\xi$ is the longitudinal correlation length with $0 \leq \xi \leq 2 R_l$ and $R_l=R/(\sqrt{2}\gamma_\mathrm{beam})$ is the boosted WS radius. A similar model was used for single nucleons in \cite{Matsuda:2023gle}. The $R_l$ dependent factor in Eq.~\eqref{eq:Uxi} guarantees consistency with two limiting cases w.r.t.~the longitudinal correlation length $\xi$.

In the limit $\xi\rightarrow 0$, Eq.~\ref{eq:Uxi} becomes
\begin{align}
    \lim_{\xi \rightarrow 0} U_{\xi}(x^\pm - y^\pm) = \delta(x^\pm - y^\pm).
\end{align}
and thus the correlator in Eq.~\eqref{eq:correlator_factorized} can be written as
\begin{align}
\label{eq:3d_mv}
    &\langle \rho^a(x^\pm,\xperp) \rho^b(y^\pm,\yperp)\rangle\nn\\
    &\quad\quad=g^2\mu^2\delta^{ab}T(x^\pm, \xperp)\delta(x^\pm-y^\pm)\delta^{(2)}(\xperp-\yperp).
\end{align}
In this limit, our model is similar to the traditional MV model with longitudinal extent (see e.g.~\cite{Iancu:2002aq} or \cite{Fukushima:2007ki}).
Introducing a transverse charge density
\begin{align}
    \rho_\perp^a(\xperp) = \intop_{-\infty}^{+\infty}dx^{\pm}\, \rho^a(x^\pm,\xperp),
\end{align}
we can integrate out the longitudinal dependence in Eq.~\eqref{eq:3d_mv} and obtain
\begin{align}
    \langle \rho_\perp^a(\xperp) \rho_\perp^b(\yperp)\rangle = g^2\mu^2\delta^{ab}\delta^{(2)}(\xperp\!-\! \yperp) \intop_{-\infty}^{+\infty}d{x^\pm}\,T(x^\pm, \xperp).
\end{align}
The normalization in Eq.~\eqref{eq:T_norm} now ensures that we recover the two-dimensional MV model \cite{McLerran:1993ka, McLerran:1993ni} at the center of our nuclei, $\xperp = \zeroperp$. We refer to this limit as the MV model limit.

On the other hand, we may take $\xi \rightarrow 2 R_l$ and obtain
\begin{align}
    \lim_{\xi \rightarrow 2 R_l} U_{\xi}(x^\pm - y^\pm) = \frac{1}{\sqrt{8 \pi R_l^2}} = \mathrm{const}.
\end{align}
In this case, the charge densities can be written as
\begin{align}
\rho^a(x^\pm, \xperp) \propto g \mu \sqrt{T(x^\pm, \xperp)} \chi_\perp^a(\xperp), 
\end{align}
where the two-dimensional Gaussian random field $\chi_\perp$ is given by
\begin{align}
    \langle \chi^a_\perp(\xperp) \rangle = 0, \quad \langle \chi^a_\perp(\xperp) \chi^b_\perp(\yperp) \rangle = \delta^{ab} \delta^{(2)}(\xperp - \yperp).
\end{align}
The charge density does not fluctuate along the longitudinal coordinate and
is merely modulated by the enveloping profile $\sqrt{T(x^\pm, \xperp)}$. This is the coherent limit of our model. Initial conditions with these particular longitudinal correlations have been previously studied using (3+1)D lattice simulations \cite{Gelfand:2016yho, Ipp:2017lho}, albeit without the non-trivial transverse structure imposed by $T(x^\pm, \xperp)$.

To sample charge distributions from our nuclear model for arbitrary $\xi$, we first generate random Gaussian noise $\chi^a(x^\pm, \xperp)$, which fulfills
\begin{align}
    \langle \chi^a(x^\pm, \xperp)\rangle &= 0,\\
    \langle \chi^a(x^\pm, \xperp)\chi^b(y^\pm, \yperp)\rangle &= \delta^{ab}\delta(x^\pm - y^\pm) \delta^{(2)}(\xperp - \yperp).
\end{align}
We then move to Fourier space with respect to $x^\pm$ and introduce a new field
\begin{align}
    \tilde \zeta^a (k^\pm, \xperp) = \sqrt{\tilde U_\xi(k^\pm)}\tilde \chi^a(k^\pm, \xperp).
\end{align}
After transforming back to position space, we obtain the single nucleus color charge densities as
\begin{align}
    \rho^a(x^\pm,\xperp) = g\mu\sqrt{T(x^\pm,\xperp)}\zeta^a(x^\pm, \xperp). \label{eq:color_charges}
\end{align}
Note that the individual color components $\rho^a$ are statistically independent. Once we have sampled charge densities $\rho_{A/B}$ for both nuclei, we obtain the corresponding single nucleus color fields by solving the transverse Poisson equations \eqref{eq:A_poisson} and \eqref{eq:B_poisson} in momentum space. We obtain
\begin{align}
    \phi^a_{A/B}(x^\pm, \xperp) = \intop_{\kperp}\frac{\tilde \rho^a_{A/B} (x^\pm, \kperp)}{\kperp^2+m^2}e^{-\kperp^2/(2\Lambda_\mathrm{UV}^2)}e^{-\ii\kperp\cdot\xperp},
\end{align}
where we have introduced an infrared (IR) cutoff $m$ and an ultraviolet (UV) cutoff $\Lambda_\mathrm{UV}$. In practice, we sample the nuclear model on a discrete lattice. The sampling procedure can be carried out numerically using the Fast Fourier Transform (FFT) algorithm, see e.g.~\cite{Schlichting:2020wrv, Matsuda:2023gle}.

\begin{figure}
    \centering
    \includegraphics{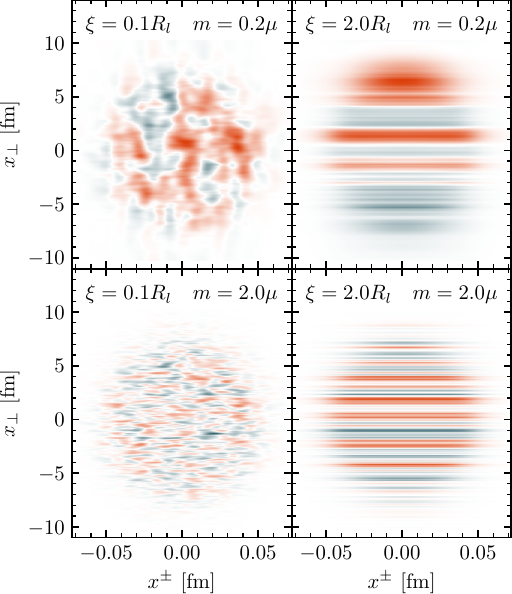}
    \caption{Slices through the center of a single component of a nucleus' color field for different values of infrared cutoff $m$ and longitudinal correlation length $\xi$. Positive (negative) values are shown in orange (blue).}
    \label{fig:single_nucleus}
\end{figure}

In Fig.~\ref{fig:single_nucleus} we show one component of the color field $\phi(x^\pm, \xperp)$ of a single nucleus sampled from our nuclear model. We cut along the center of one transverse direction and compare different values of the IR cutoff $m$ and longitudinal correlation length $\xi$, showing positive values in orange and negative values in blue. While the magnitude of $m$ governs the size of transverse fluctuations, $\xi$ determines the longitudinal structure of our nuclei. The right panels show the coherent limit $\xi = 2 R_l$.

\section{Numerical results for the (3+1)D structure of the Glasma}%
\label{sec:results}

We now discuss the (3+1)D spacetime structure of the Glasma emerging from our model. We compute the integrals in Eqs.~\eqref{eq:f+-}--\eqref{eq:fij} using the Monte Carlo approach outlined in Sec.~\ref{sec:mc_sampling}.
From the Monte Carlo integration, we obtain estimates for the Glasma field strength tensor $f^{\mu\nu}$ and energy-momentum tensor $T^{\mu\nu}$ given in Eq.~\eqref{eq:EMT}.
The raw data files for $T^{\mu\nu}$ are published in~\cite{zenodo:RHIC,zenodo:LHC}.
We solve the  Landau condition Eq.~\eqref{eq:landau} 
to find the local rest frame energy density $\elrf$ and flow velocity $u^\mu$. We obtain estimates for the magnitude of the Monte Carlo errors by performing standard jackknife analysis and we choose the number of Monte Carlo samples large enough to make them negligible. In practice, we found that $M = 10^5$ samples are sufficient.

\begin{table}
  \centering
    \begin{ruledtabular}
    \begin{tabular}{llll}
    \textbf{Param.} & \textbf{Name} & \textbf{Value(s)} & \textbf{Unit} \\
    \midrule
    $N_c$ & No. of colors & 3 & - \\
    $\gamma_\mathrm{beam}$ & Lorentz factor & 100 (R), 2700 (L) & - \\
    $\sqrt{s_{NN}}$ & c.m. energy\footnote{Assuming a nucleon mass of $m_0 \approx 1 \, \mathrm{GeV}$.} & 200 (R), 5400 (L) & GeV \\
    $R$   & WS radius & 6.38 (R), 6.62 (L) & fm \\
    $d$   & WS skin depth & 0.535 (R), 0.546 (L) & fm \\
    $g$ & YM coupling & 1 & - \\
    $\mu$ & MV scale & 1 & GeV \\
    $m$   & IR cutoff & 0.2, 2.0 & GeV \\
    $\Lambda_\mathrm{UV}$ & UV cutoff & 10 & GeV \\
    $\xi$ & correlation length & 0.1, 0.5, 2.0 & $R_l$ \\
    $b$   & impact parameter & 0, 1  & $R$ \\
    $\tau$ & proper time & 0.2, 0.4, 0.6, 0.8, 1.0 & fm/c \\
    \end{tabular}
    \end{ruledtabular}
\caption{Physical model parameters and their values in our calculations with R (L) denoting RHIC (LHC) setups. The correlation length $\xi$ and the impact parameter $b$ are given as multiples of the Woods-Saxon radius and are therefore different for RHIC and LHC setups. Without loss of generality, we always put the impact parameter in transverse $x$-direction.}
\label{tab:params}
\end{table}

Table~\ref{tab:params} shows the physical parameters of our calculations and the values they are allowed to take.
We consider two different collision energies, comparable to Pb+Pb collisions at LHC and Au+Au collisions at RHIC\@. The choice of collision energy enters our computations through the beam Lorentz contraction factor $\gamma_\mathrm{beam}$. Furthermore, we adapt the parameters of our Woods-Saxon model to the different nuclei in the same way as \cite{Schenke:2012hg}.
The values for the correlation length $\xi$ correspond to longitudinal fluctuations at the scale of the nucleon size, $\xi=0.1\, R_l$, the coherent limit, $\xi=2\, R_l$, and an intermediate value, $\xi=0.5\, R_l$.
We use $N_\mathrm{ev} = 10$ independent collision events to compute event averages of our observables. As shown in Fig.~\ref{fig:shift}, we use a collision origin shifted by $\delta t = (R + d) / \gamma_\mathrm{beam}$ to avoid contributions from the background fields at larger rapidities.

We note that the dilute limit differs from the non-perturbative Glasma in one key aspect: due to the perturbative expansion, we find that the MV scale parameter $\mu$, or more generally the energy scale $g^2 \mu$, appears in our analytic results only as a prefactor in $f^{\mu\nu}$ or $T^{\mu\nu}$. 
The scale $g^2 \mu$ does not appear as a transverse momentum scale in the dilute limit. Instead, the transverse structure of the nuclei and the resulting Glasma is determined solely by the infrared regulator $m$. 
We thus interpret $m$ as the analog of the saturation momentum $Q_s$ in the non-perturbative Glasma, and $g^2 \mu$ as an energy scale that requires calibration using either experimental results or a fit to a non-perturbative lattice simulation. For example, using the dilute Glasma as an initial stage for a full simulation of a heavy-ion collision,
the parameter $g^2\mu$ may be fixed such that the charged particle multiplicity at mid-rapidity is correctly reproduced at either RHIC or LHC\@. In the present work, we simply set $g^2 \mu = 1 \, \mathrm{GeV}$, keeping in mind that phenomenological applications require a properly chosen value.

\begin{figure}
    \centering
    \includegraphics[width=\columnwidth]{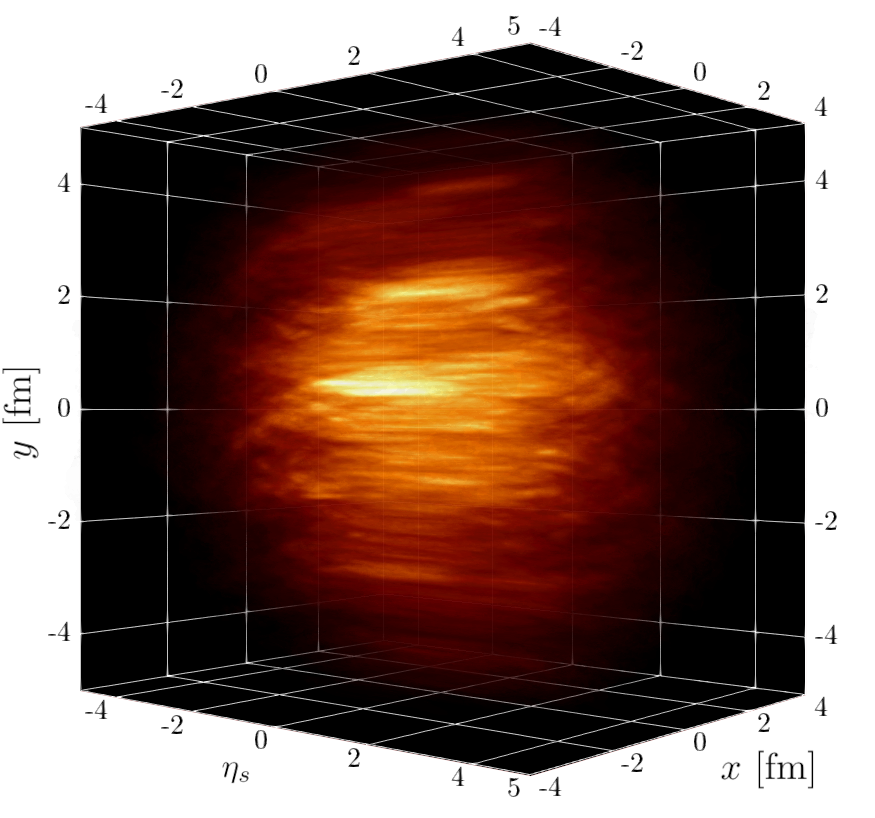}
    \caption{Perspective view of the local rest frame energy density $\elrf$ of a single collision event at $\tau=0.4\,\mathrm{fm}/c$ with impact parameter $b=R$. RHIC nuclear parameters with $\xi = 0.5\,R_l$, $m=0.2\,\mu$ are used. Lighter colors correlate with larger $\elrf$. A movie showing different viewing angles is included in the Supplemental Material~\cite{supplement:video}.}
    \label{fig:3D-elrf}
\end{figure}

\begin{figure*}
    \centering
    \includegraphics{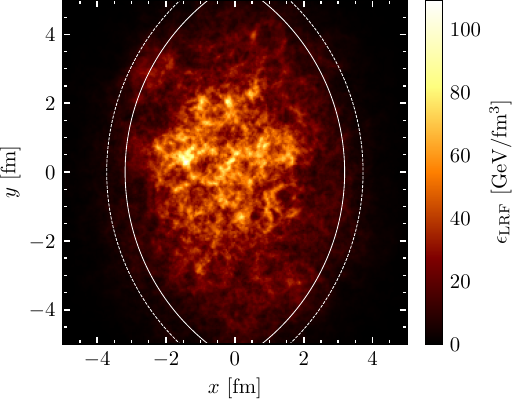}\hfill%
    \includegraphics{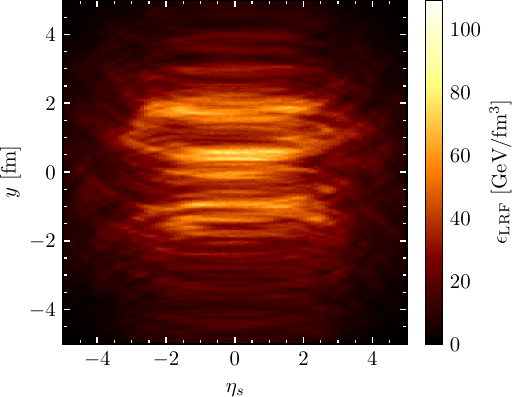}
    \caption{Transverse (left) and longitudinal (right) slices through the local rest frame energy density $\elrf$ of a single collision event at $\tau=0.4\,\mathrm{fm}/c$ with impact parameter $b=R$. RHIC nuclear parameters with $\xi = 0.5\,R_l$, $m=0.2\,\mu$ are used. The white lines represent $R$ and $R+d$ boundaries of both nuclei.}
    \label{fig:high_res_slice}
\end{figure*}

Figure~\ref{fig:3D-elrf} shows a perspective plot of the three-dimensional local rest frame energy density $\elrf$ for a single Au+Au collision event with an impact parameter $b=R$ at RHIC energy at $\tau=0.4~\mathrm{fm}/c$.
We show transverse and longitudinal slices of the energy distribution in Fig.~\ref{fig:high_res_slice}.
In the left panel, we observe the typical almond shape induced by the non-zero impact parameter.
The right panel depicts elongated, approximately boost-invariant structures with varying transverse extents reminiscent of Glasma color flux tubes \cite{Lappi:2006fp, Fujii:2008dd}. These structures are analogous to the flux tube structure employed in several initial state models for describing longitudinal correlations in the initial stage of heavy-ion collisions \cite{Pang:2015zrq, Broniowski:2016xnz}. 

\subsection{Longitudinal structure and flow}%
\label{sec:longidudinal_flow}

The three-dimensional Glasma generated in collisions of longitudinally extended nuclei has a particular longitudinal structure and flow properties that are qualitatively different from the boost-invariant scenario. For practical reasons, we compute the energy-momentum tensor in a (shifted) Milne frame, parameterized by proper time $\tau$ and spacetime rapidity $\eta_s$, as discussed in Sec.~\ref{sec:emt_theory}.  We find that the longitudinal flow $u^\eta$ of the collision medium, i.e.~the rapidity component of the four-velocity $u^\mu$, deviates from the idealized Bjorken case, particularly at large rapidities.
This is in contrast to the boost-invariant Glasma, where $u^\eta$ exhibits only local fluctuations \cite{McDonald:2017eml} and thus $T^{\tau\tau}$ can be considered as a reasonable measure for the energy density of the Glasma. 
The Milne frame centered at the collision point is naturally adapted to the symmetries of the boost-invariant Glasma, which (assuming negligible dynamics in the transverse direction) expands with $u^\tau=1$ and $u^\eta=0$, known as free streaming. As argued in Sec.~\ref{sec:emt_theory}, there is no unique Milne frame for a collision of longitudinally extended nuclei. The shifted origin of the Milne frame compared to the collision center, along with the extended collision region leads to a violation of the free streaming behavior. The three-dimensional Glasma picks up a considerable longitudinal velocity $u^\eta$ in regions of large rapidity.
To avoid ambiguity due to the choice of frame, it is therefore sensible to study frame-independent quantities such as the LRF energy density $\elrf$, or the transverse pressure $T^{xx} + T^{yy}$, which is unaffected by longitudinal boosts.

\begin{figure}
    \centering\includegraphics{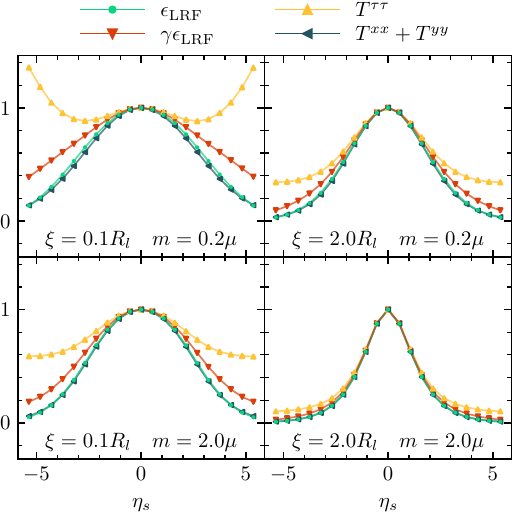}
    \caption{Different formulations of energy density integrated over the transverse plane as a function of spacetime rapidity $\eta_s$ at fixed proper time $\tau=0.4\,\mathrm{fm}/c$.
    The results are normalized at $\eta_s =0$ and are averages over 10 central collision events at RHIC energy.
    The local rest frame energy density $\elrf$ (emerald dots) agrees with the transverse pressure $T^{xx} + T^{yy}$ (blue triangles).
    For small $\xi$ and $m$ (upper left) the $T^{\tau\tau}$ component increases for large $\eta_s$.}
    \label{fig:energy-comp}
\end{figure}

In Fig.~\ref{fig:energy-comp} we investigate the rapidity dependence of various notions of energy density in the dilute Glasma. We see that the LRF energy density $\elrf$ is in almost perfect agreement with the sum of transverse pressures $T^{xx}+T^{yy}$. However, note that the $T^{\tau\tau}$ profiles differ fundamentally from the $\elrf$ profiles at large $\eta_s$. In addition to the aforementioned energy densities and pressures, we also show $\gamma \elrf$, which is the LRF energy density, weighted by the local flow component $\gamma(\tau, \eta_s, \xperp) \equiv u^\tau(\tau, \eta_s, \xperp)$. This quantity is interesting because $\gamma \elrf$ is used to determine the transverse geometric structure of the collision medium in terms of eccentricity coefficients (see Sec.~\ref{sec:eccentricity}). We find that its rapidity profile is typically wider, but similarly well-behaved at large rapidities as $\elrf$ and transverse pressure.

\begin{figure}
    \centering\includegraphics{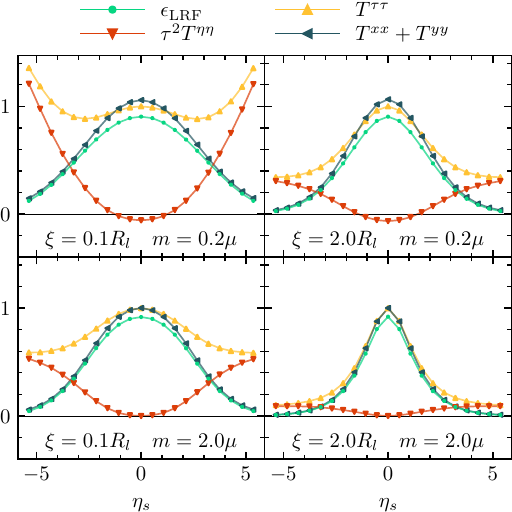}
    \caption{Comparison of the components of the energy-momentum tensor $T^{\mu\nu}$ and the local rest frame energy density $\elrf$ integrated over the transverse plane as a function of spacetime rapidity $\eta_s$ at fixed proper time $\tau=0.4\,\mathrm{fm}/c$. The results are normalized to $T^{\tau\tau}(\eta_s = 0)$ and are averages over 10 central collision events at RHIC energy. Tracelessness of $T^{\mu\nu}$ is manifest with the longitudinal pressure $\tau^2T^{\eta\eta}$ (red triangles) rising and compensating for the large $T^{\tau\tau}$ (yellow triangles) at extreme $\eta_s$. For small $m$ (top row) $\tau^2T^{\eta\eta}$ is negative at central $\eta_s$.}
    \label{fig:Ttautau}
\end{figure}

Figure \ref{fig:energy-comp} reveals a considerable difference between $T^{\tau\tau}$,
which is the energy density reported by an observer moving with $u^\tau=1$ and $u^\eta=0$,
and $\elrf$ at large rapidities. Intuitively, this means that the flow of the three-dimensional Glasma differs from longitudinal Bjorken flow, which makes the Milne frame an ill-adapted coordinate system. 
Consequently, the Milne frame energy-momentum tensor $T^{\mu\nu}$ picks up sizeable off-diagonal components at large rapidities. 
Upon diagonalization and consequential transformation to the LRF, $\elrf$ becomes much smaller than $T^{\tau\tau}$ and we find significant contributions to the local $u^\eta$.
In addition to longitudinal flow, the mismatch $T^{\tau\tau} \gg \elrf$ also arises from longitudinal pressure, 
as shown in Fig.~\ref{fig:Ttautau}. Particularly for small $\xi$ and $m$, we find that the longitudinal pressure $\tau^2 T^{\eta\eta}$ dominates over the transverse pressure at large rapidities. This is consistent with the fact that the energy-momentum tensor is traceless. Similar to $T^{\tau\tau}$, it is apparent that longitudinal pressure is strongly affected by the choice of coordinate system.

\begin{figure}
    \centering
    \includegraphics{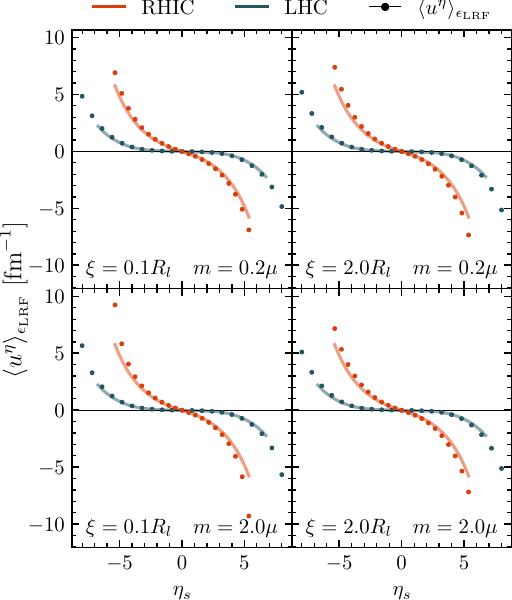}
    \caption{Longitudinal flow $u^\eta$ weighted with local rest frame energy density and integrated over the transverse plane as a function of spacetime rapidity $\eta_s$ at $\tau=0.4\,\mathrm{fm}/c$. Discrete points represent the average of 10 central collision events. Lines are analytic predictions from simple model considerations (see Appendix \ref{app:ueta}).
    }
    \label{fig:ueta_curves}
\end{figure}

We study longitudinal flow by computing an $\elrf$-weighted average of $u^\eta$ in the transverse plane, i.e.
\begin{align}
    \langle u^\eta \rangle_{\elrf} = \left\langle \frac{\int_{\xperp}u^\eta (\tau, \eta_s, \xperp)\, \elrf(\tau, \eta_s, \xperp)}{\int_{\xperp} \elrf(\tau, \eta_s, \xperp)} \right\rangle_\mathrm{events}.
\end{align}
In Fig.~\ref{fig:ueta_curves} we show this weighted average
over $\eta_s$ for constant $\tau=0.4\,\mathrm{fm}/c$. Regions of positive $\eta_s$ correspond to negative $u^\eta$ and vice versa. This indicates a longitudinal expansion that is slower than Bjorken flow, to which the Milne frame is naturally adapted, and leads to the strong deviations of $T^{\tau\tau}$ from the rest frame energy density $\elrf$.
We find that the shape of the longitudinal flow can be explained in a simple model, depicted by the solid curves, where we assume that the resulting four-velocity at a given point in the forward light cone can be obtained by a superposition of Bjorken flows starting from each point in the collision region taking into account our forward-shifted coordinate system. This model is derived in Appendix~\ref{app:ueta}. Since the data fit the analytic results extraordinarily well, we conclude that the longitudinal flow can mostly be attributed to the extended collision region and the shifted origin of our coordinate system. This reinforces the role of $\elrf$ and $T^{xx} + T^{yy}$ as the more fundamental and physical notions of energy density in the three-dimensional Glasma, as compared to the Milne frame energy density $T^{\tau\tau}$.

\subsection{Time and energy dependence}

\begin{figure}
    \centering
    \includegraphics{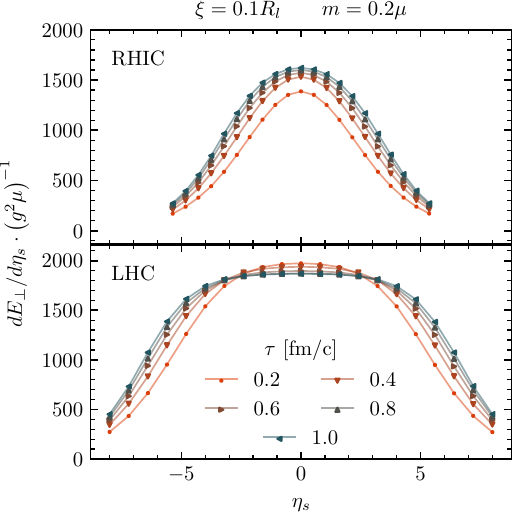}
    \caption{Differential transverse energy $dE_\perp/d\eta_s$ as a function of spacetime rapidity $\eta_s$ for different values of proper time $\tau$. The points represent the average of 10 central collision events. The rapidity profiles stabilize at late times $\tau \gtrsim 0.6 \, \mathrm{fm}/c$. For collisions at LHC energy, we observe a plateau around mid-rapidity, which is absent at RHIC energy.}
    \label{fig:elrf_tau}
\end{figure}

In addition to the rapidity dependence of the energy density, we also study its time and collision energy dependence. 
In Fig.~\ref{fig:elrf_tau} we show the rapidity profiles of the transverse energy per unit spacetime rapidity $dE_\perp/d\eta_s$ for different values of proper time $\tau$. These data are obtained by integrating the sum of transverse pressures $T^{xx} + T^{yy}$ over the transverse plane and multiplying with $\tau$ to correct for the expected leading order time dependence:
\begin{align}
    \frac{dE_\perp}{d\eta_s} = \tau \intop_\xperp\big(T^{xx}(\tau, \eta_s, \xperp)+ T^{yy}(\tau, \eta_s, \xperp)\big).
\end{align}
We see that the transverse energy per unit rapidity stabilizes on a time scale of $\sim 1.0\, \mathrm{fm}/c$. 
This is a well-known result for the boost-invariant case~\cite{Lappi:2003bi, Lappi:2006hq}, where the energy density exhibits a characteristic $1/\tau$ behavior associated with free streaming for $\tau \gtrsim 0.2\, \mathrm{fm}/c$. We find that this holds for the three-dimensional Glasma as well, even at extreme rapidities.
For LHC energy we observe a stable plateau of roughly $\pm 3.5$ units in rapidity around $\eta_s =0$ which can be identified with the boost-invariant regime.
This plateau is absent at RHIC energy, where the profiles appear to be approximately Gaussian. 
A similar plateau for LHC energy appears in the curves of the longitudinal flow in Fig.~\ref{fig:ueta_curves}.

\subsection{Limiting fragmentation}

\begin{figure}
    \centering\includegraphics{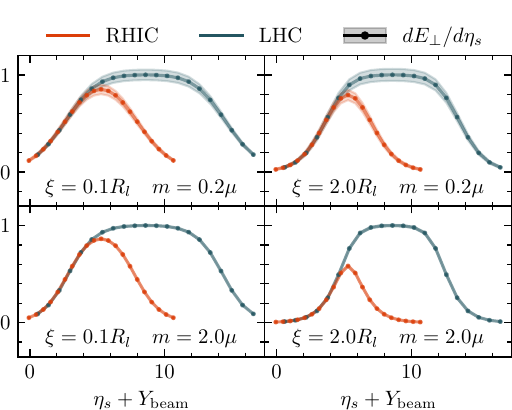}
    \caption{Differential transverse energy as a function of spacetime rapidity $\eta_s$ at fixed proper time $\tau=0.4\,\mathrm{fm}/c$. The curves are normalized to the respective LHC values at $\eta_s=0$ and are offset by the respective LHC and RHIC beam rapidities $Y_\mathrm{beam}$. The results are averaged over 10 central collision events with error bands for one standard deviation. We find that the flanks of the rapidity profiles coincide irrespective of the collision energy, known as limiting fragmentation.}
    \label{fig:limiting_fragmentation}
\end{figure}

In Fig.~\ref{fig:limiting_fragmentation} we investigate rapidity profiles of the differential transverse energy $dE_\perp/d\eta_s$ for different collision energies $\sqrt{s_{NN}}$. After shifting the profiles by the respective beam rapidity $Y_\mathrm{beam}=\arcosh(\gamma_\mathrm{beam})$, we see remarkable agreement between LHC and RHIC setups regarding the shape of the flanks at large rapidities. In other words, for extreme rapidities (the fragmentation region) the rapidity distribution of the transverse energy is independent of the collision energy. This is an indication of limiting fragmentation, first observed experimentally for charged particle distributions in the context of proton-antiproton collisions \cite{UA5:1986yef}, after a theoretical description was given in \cite{Benecke:1969sh}. It was subsequently studied experimentally e.g. for p+A collisions \cite{Elias:1979cp}, d+Au collisions \cite{PHOBOS:2004fzb} and Au+Au collisions at RHIC \cite{BRAHMS:2001llo, Back:2002wb}. From our results, we expect limiting fragmentation to hold for LHC energies as well. We can confirm limiting fragmentation also for the local rest frame energy density $\elrf$ as well as $T^{\tau\tau}$. Therefore, we conclude that it is truly a universal effect that governs the overall scaling behavior of the energy-momentum tensor in the dilute approximation.

\subsection{Eccentricity}\label{sec:eccentricity}

Up until now, we have highlighted the longitudinal structure of the three-dimensional Glasma, but our approach allows us to study the transverse structure as well, namely in the form of (transverse) eccentricity coefficients. 
From the local rest frame energy density $\elrf(\tau, \eta_s, \xperp)$ we obtain the $n$\nobreakdash-th eccentricity via the standard formula
\begin{align}
\label{eq:eps-ecc}
\varepsilon_n(\tau,\eta_s)= 
    \frac{
        \int_\xperp \gamma\ \elrf\ \bar{r}^n \exp(\ii n\bar\phi)
        }{
        \int_\xperp \gamma\ \elrf\ \bar r^n
    }  
\end{align}
with $\elrf = \elrf(\tau, \eta_s, \xperp)$ and $\gamma = \gamma(\tau, \eta_s, \xperp) = u^\tau(\tau, \eta_s, \xperp)$ the local Lorentz factor. In the above, $\bar r=\sqrt{(x-x_0)^2 + (y-y_0)^2}$ and $\bar \phi=\arctan\big((y-y_0)/(x-x_0)\big)$ are polar coordinates in the center of mass frame in the transverse plane. We evaluate the transverse center of mass,
\begin{align}
    \xperp_0 &= \frac{\int_\xperp \xperp\, \elrf(\xperp) \gamma(\xperp)}{\int_\xperp \elrf(\xperp) \gamma(\xperp)},
\end{align}
at mid-rapidity, $\eta_s = 0$.

\begin{figure}
    \centering
    \includegraphics{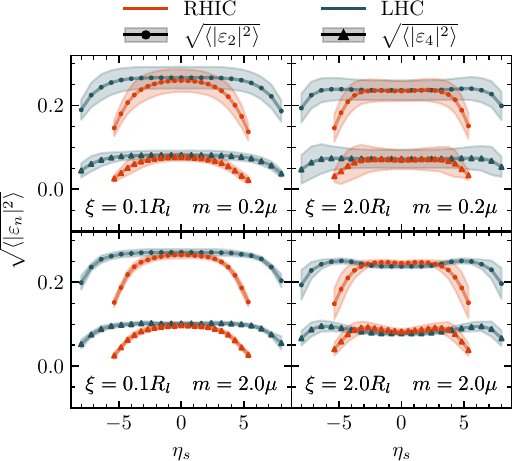}
    \caption{Eccentricities $\sqrt{\langle|\varepsilon_2|^2\rangle}$ and $\sqrt{\langle|\varepsilon_4|^2\rangle}$ at $\tau=0.4\,\mathrm{fm}/c$ as a function of spacetime rapidity $\eta_s$. The impact parameter $b=R$ and $\langle\cdot\rangle$ refers to an average over 10 events with the error bands representing one standard deviation in the event-by-event statistics.}
    \label{fig:ecc}
\end{figure}

In Fig.~\ref{fig:ecc} we show the rapidity dependence of the eccentricities $\sqrt{\langle|\varepsilon_2|^2\rangle}$ and $\sqrt{\langle|\varepsilon_4|^2\rangle}$ for both RHIC and LHC energies and different combinations of $\xi$ and $m$. We choose a large impact parameter of $b=R$ and consequently see significant contributions to $\varepsilon_2$ and $\varepsilon_4$. The eccentricities are independent of rapidity except for the most forward and backward bins, where they fall off. We note that the central plateau is narrower for RHIC than for the LHC setup. The fourth eccentricity $\varepsilon_4$ exhibits similar qualitative behavior to the second eccentricity $\varepsilon_2$ but has a lower overall value. In any case, $\varepsilon_2$ and $\varepsilon_4$ show only minor dependence on the longitudinal correlation length $\xi$.

Evidently, it would also be interesting to study the longitudinal decorrelation of the event geometry in the dilute Glasma by considering un-equal rapidity correlations of the eccentricities~(see e.g. \cite{Schenke:2016ksl,Garcia-Montero:2023gex,Schenke:2022mjv}). However, the fluctuations of the eccentricities $\varepsilon_{n}(\eta_s)$ are largely driven by fluctuations in nucleon positions, which are currently not included in our model and we therefore defer this analysis to a forthcoming study.

\begin{figure}
    \centering\includegraphics{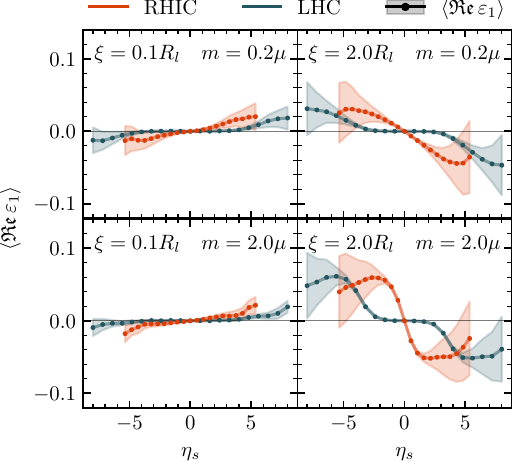}
    \caption{Real part of the eccentricity $\varepsilon_1$ at $\tau=0.4 \, \mathrm{fm}/c$  for an impact parameter $b=R$. The points represent the average of 10 events and the error bands represent the standard deviation of the event-by-event statistics.}
    \label{fig:eps_ecc1}
\end{figure}

In Fig.~\ref{fig:eps_ecc1} we show the real part of the first eccentricity $\varepsilon_1$ obtained from Eq.~\eqref{eq:eps-ecc} for $n=1$. Since the impact parameter lies in the $x$-direction we only get negligible contributions to the imaginary part. Figure~\ref{fig:eps_ecc1} shows how the Glasma is skewed in the transverse plane. Specifically, for longitudinal fluctuations at the scale of the nucleon size, $\xi=0.1\, R_l$, regions of positive rapidity feature more energy density for positive $x$. This is what one would expect from the collision geometry in the sense that nucleus $B$ is shifted to positive $x$ and moves in positive $z$-direction. 
Interestingly, in the coherent limit $\xi=2.0\, R_l$, the opposite is the case and we see more energy in regions of positive $x$ for negative rapidity. 
In our results, the energy distribution appears to depend strongly on the longitudinal structure of the nuclei---even changing signs for different $\xi$.

Generically, we find that the dilute (3+1)D Glasma exhibits both non-trivial longitudinal flow $\langle u^\eta \rangle \neq 0$ and a skewed energy distribution in the transverse plane, $\varepsilon_1 \neq 0$. In \cite{Ryu:2021lnx} it was demonstrated, using a parameterized Glauber model, that longitudinal flow and skewness may both contribute to the directed flow of pions in off-central Au+Au collisions. We find that these features also arise naturally in the dilute (3+1)D Glasma. 
Thus, an interesting prospect would be to study the impact of the longitudinal structure on observables such as directed flow.

Finally, we remark that the RHIC and LHC curves' shapes are similar for large rapidities, but around mid-rapidity, a plateau becomes visible for LHC that is absent in the RHIC curves.

\section{Conclusion}\label{sec:conclusion}

We presented the semi-analytical treatment of the three-dimensional, dilute Glasma in the weak field approximation. Starting from the expressions for color fields obtained in \cite{Ipp:2021lwz} within the dilute limit of the Color Glass Condensate effective field theory, we derived concise expressions for the field strength tensor and energy-momentum tensor of the dilute Glasma. We found that our analytical results have a transparent geometric interpretation in terms of integrating over the causal past of the specific spacetime point at which we evaluate the energy-momentum tensor.

The main advantage of the dilute approximation is that it leads to analytic results for the Glasma field strength tensor. The corresponding integrals have to be solved numerically, but, compared to non-perturbative real-time (3+1)D lattice simulations~\cite{Ipp:2020igo, Schlichting:2020wrv, Matsuda:2023gle, Matsuda:2024moa},
the numerical evaluation using Monte Carlo methods can be implemented much more efficiently than previous simulations at comparable spacetime volumes and enables system sizes that were not achievable previously. As such, we are able to describe in unprecedented detail the energy-momentum tensor and derived observables of the three-dimensional Glasma, accounting for both transverse and longitudinal structure within the colliding nuclei.
Furthermore, our calculation of the field strength tensor does not require numerical time-stepping common to traditional simulation approaches. 
Instead, the field strengths can be evaluated at arbitrary points in the future light cone. This allows the direct evaluation of the energy-momentum tensor at arbitrary hypersurfaces and simplifies the coupling to hydrodynamics or kinetic theory.
Especially for situations where many independent events have to be considered to obtain a given observable, we believe that the dilute Glasma offers a novel and economic approach to the pre-equilibrium stage of relativistic heavy-ion collisions.

By introducing a generalized McLerran-Venugopalan model with a fully three-dimensional structure for the color charge distributions of our nuclei we were able to parametrize longitudinal correlations within the nuclei.
This led to a longitudinally extended overlap region for symmetric nuclear collisions and required us to shift the Milne frame origin into the future light cone outside of the overlap region.
Within this setup, we explored the longitudinal and transverse structures of Pb+Pb and Au+Au collisions at LHC and RHIC energies.
We carefully discussed how to interpret the Milne frame components of the energy-momentum tensor of the resulting Glasma.
We found that the energy distribution of the Glasma exhibits broken boost-invariance and non-trivial longitudinal flow at large rapidities.
Remarkably, we recovered limiting fragmentation as a generic feature of the dilute approximation.
We believe that this can be demonstrated analytically, as will be detailed in a forthcoming manuscript.
We also studied the pre-equilibrium, transverse structure in terms of eccentricity coefficients $\varepsilon_n$.
In particular, we find that off-central collisions lead to a rapidity-odd first-order eccentricity coefficient $\varepsilon_1$.

There are multiple possible extensions to this work.
As a direct application of our approach, the dilute Glasma can be used as fully (3+1)D initial conditions for effective kinetic theory~\cite{Arnold:2002zm, Kurkela:2015qoa, Keegan:2016cpi, Kurkela:2018vqr} or hydrodynamic simulations of the QGP~\cite{Schenke:2010rr, Gale:2013da, McDonald:2020oyf, McDonald:2023qwc}.
Proceeding through these later stages of heavy-ion collisions would allow us to compute observables directly comparable to those measured in experiments and study their dependence on our fully three-dimensional nuclear model.
Furthermore, we are actively working on improving our nuclear model, for example by considering nucleons~\cite{Matsuda:2023gle} or nucleon hot spots~\cite{Schlichting:2014ipa, Mantysaari:2016ykx, Demirci:2021kya} as additional substructures within the nuclei.
We are excited to explore the implications of such improved models on the various different aspects of the (3+1)D spacetime structure of the initial state of high-energy heavy-ion collisions in the future.

\begin{acknowledgments}
\hyphenation{Deut-sche}
\hyphenation{For-schungs-ge-mein-schaft}
\hyphenation{Dok-to-rats-kol-leg}

ML, DM and KS are supported by the Austrian Science Fund FWF No.\ P34764\@. ML and KS further acknowledge funding from the Doktoratskolleg Particles and Interactions (DK-PI, FWF doctoral program No.\ W1252)\@.
SS acknowledges support by the Deutsche Forschungsgemeinschaft (DFG, German Research Foundation) through the CRC-TR 211 ’Strong-interaction matter under extreme conditions’ – project number 315477589 – TRR 211\@.
PS is supported by the Academy of Finland, by the Centre of Excellence in
Quark Matter (project 346324), and European Research Council, ERC-2018-ADG-835105 YoctoLHC\@.
The computational results presented have been achieved in part using the Vienna Scientific Cluster (VSC).
\end{acknowledgments}

\appendix

\section{Derivation of the dilute Glasma field strength tensor}
\label{sec:derivation}

To compute the dilute Glasma field strength tensor given in Eq.~\eqref{eq:dilute_f} we require first derivatives of the Glasma gauge fields $a^\mu(x)$. For the $f^{+-}$ component, we start by working out $\p^+ a^- = \p_-a^-$
\begin{align}
    \p_-a^- &= \frac{g}{2} f_{abc} t^c \intop_{\pperp ,\qperp}\intop_0^{\infty} dv^+ \intop_0^{\infty} dv^-\, \tilde \phi^a_A( x^+\! - v^+,\pperp) \nonumber \\
    &\ind\times \p_-^{(x)}\tilde \phi^b_B(x^-\! - v^-, \qperp)\frac{\big((\pperp\! +\! \qperp)^2 - 2 \pperp^2\big) v^-}{ |\pperp\! +\! \qperp | \tau' } \nonumber \\
    &\ind\times J_1( |\pperp\! +\! \qperp | \tau') e^{-\ii(\pperp+ \qperp)\cdot \xperp},
\end{align}
which can be simplified by first rewriting the $\p_-^{(x)}$ as $-\p_-^{(v)}$. 
Integrating by parts  then gives
\begin{align}
    \p^+ a^-&=\frac{g}{2} f_{abc} t^c \intop_{\pperp ,\qperp}\intop_0^{\infty} dv^+ \intop_0^{\infty} dv^- \tilde \phi^a_A( x^+\! - v^+,\pperp)  \nonumber \\
    &\ind\times \tilde \phi^b_B(x^-\! - v^-, \qperp)\big((\pperp\! +\! \qperp)^2 - 2 \pperp^2\big) \nn \\
 &\ind\times\frac{J_0( |\pperp\! +\! \qperp | \tau')}{2} e^{-\ii(\pperp+ \qperp)\cdot \xperp}.
\end{align}
We dropped the boundary terms because $\tilde\phi^b_B(x^-\!-v^-,\qperp)$ is zero for finite $x^-$ if $v^-\rightarrow \infty$ and goes to zero if $v^-=0$ and $x^-$ is sufficiently far away from the tracks of the nuclei, which we already assumed. In a similar fashion,
\begin{align}
    -\p^- a^+&= \frac{g}{2} f_{abc} t^c\!\intop_{\pperp ,\qperp}\!\intop_0^{\infty} dv^+ \!\intop_0^{\infty}\! dv^-\, \tilde \phi^a_A( x^+\! - v^+,\pperp) \nonumber \\
    &\ind \times  \tilde \phi^b_B(x^-\! - v^-, \qperp) \big((\pperp\! +\! \qperp)^2 - 2 \qperp^2 \big)\nn\\
    &\ind\times \frac{J_0( |\pperp\! +\! \qperp | \tau')}{2} e^{-\ii(\pperp+ \qperp)\cdot \xperp},
\end{align}
which gives
\begin{align}
    &f^{+-}= g f_{abc} t^c \intop_{\pperp ,\qperp}\intop_0^{\infty} dv^+ \!\intop_0^{\infty} dv^-\, \tilde \phi^a_A( x^+\! - v^+,\pperp)  \nonumber \\
    &\hspace{0.3cm}\times \tilde \phi^b_B(x^-\! - v^-, \qperp)(\pperp \cdot \qperp)~
J_0( |\pperp\! +\! \qperp | \tau') e^{-\ii(\pperp+ \qperp)\cdot \xperp}.
\end{align}
To find an expression for $f^{+i}$ we compute
\begin{align}
    \p_- a^i&=-\ii\frac{g}{2} f_{abc} t^c \intop_{\pperp ,\qperp}\intop_0^{\infty} dv^+ \intop_0^{\infty} dv^- \tilde \phi^a_A( x^+\! - v^+,\pperp) \nonumber \\ 
    &\ind\times \tilde \phi^b_B(x^-\! - v^-, \qperp)(p^i\!-\! q^i)\frac{|\pperp\! +\! \qperp|v^+}{\tau'}\nn\\
    &\ind\times J_1( |\pperp\! +\! \qperp | \tau') e^{-i(\pperp+ \qperp)\cdot \xperp},
\end{align}
which is obtained by partial integration as above. 
Furthermore, we use $-\p^i a^+ = \p_i a^+$ to get
\begin{align}
    \p_i a^+ &= -\ii\frac{g}{2} f_{abc} t^c  \intop_{\pperp,\qperp}  \intop_0^{\infty}   dv^+   \intop_0^{\infty} dv^- \tilde \phi^a_A( x^+\!  - v^+, \pperp) \nonumber \\
   &\ind\times \tilde \phi^b_B(x^-\! - v^- , \qperp) (p^i +q^i)\frac{
\big( - (\pperp\! +\! \qperp)^2 + 2 \qperp^2 \big) v^+
}{|\pperp\! +\! \qperp | \tau'} \nonumber \\
&\ind\times J_1( |\pperp\! +\! \qperp | \tau') e^{-\ii(\pperp+ \qperp)\cdot \xperp}.
\end{align}
Combining these two expressions yields
\begin{align}
    &f^{+i} =  -\ii gf_{abc} t^c  \intop_{\pperp,\qperp}  \intop_0^{\infty}   dv^+   \intop_0^{\infty} dv^- \tilde \phi^a_A( x^+\! - v^+, \pperp) \nonumber \\
    &\ind\times \tilde \phi^b_B(x^-\! - v^- , \qperp) \frac{v^+}{|\pperp\! +\! \qperp | \tau'}\left[p^i\qperp^2-q^i\big(\pperp^2+2(\pperp\cdot\qperp)\big)\right] \nonumber \\
    &\ind\times J_1( |\pperp\! +\! \qperp | \tau') e^{-i(\pperp+ \qperp)\cdot \xperp}.\label{eq:fpi}
\end{align}
Analogously, one finds
\begin{align}
    &f^{-i} =  -\ii gf_{abc} t^c  \intop_{\pperp,\qperp}  \intop_0^{\infty}   dv^+   \intop_0^{\infty} dv^- \tilde \phi^a_A( x^+\!-  v^+, \pperp) \nonumber \\
    &\ind\times \tilde \phi^b_B(x^-\! - v^- , \qperp)  \frac{v^-}{|\pperp\! +\! \qperp | \tau'}\left[p^i\big(\qperp^2+2(\pperp\cdot\qperp)\big)-q^i\pperp^2 \right] \nonumber \\
    &\ind \times J_1( |\pperp\! +\! \qperp | \tau') e^{-i(\pperp+ \qperp)\cdot \xperp}. 
\end{align}
Finally, it is straightforward to compute
\begin{align}
    &f^{ij} = -g f_{abc} t^c \intop_{\pperp ,\qperp}\intop_0^{\infty} dv^+ \intop_0^{\infty} dv^- \tilde \phi^a_A( x^+\! - v^+,\pperp) \nonumber \\
    &\ind\times \tilde \phi^b_B(x^-\! - v^-, \qperp)(q^i p^j - q^j p^i)J_0( |\pperp\! +\! \qperp | \tau') e^{-i(\pperp+ \qperp)\cdot \xperp}.
\end{align}

This provides the components of the field strength tensor expressed as six-dimensional integrals. These integrals can be computed numerically, but the performance of such a computation is greatly enhanced by reducing the number of integrals. Fortunately, we can reduce all components of the field strength tensor to three-dimensional integrals, as we will show in the following. Starting with $f^{+-}$ we introduce
\begin{align}
    \tilde \beta^{i,a}_{A}(x^+\!-v^+,\pperp) &= \ii p^i\tilde \phi^{a}_{A}(x^+\!-v^+,\pperp),\\
    \tilde \beta^{i,b}_{B}(x^-\!-v^-,\qperp) &= \ii q^i\tilde \phi^{b}_{B}(x^-\!-v^-,\qperp),
\end{align}
or, equivalently,
\begin{align}
\beta^{i,a}_A(x^+\!-v^+,\uperp) &= \partial^i_{(u)} \phi_A^a(x^+\!-v^+,\uperp),\\
\beta^{i,b}_B(x^-\!-v^-,\sperp) &= \partial^i_{(s)}\phi^b_B(x^-\!-v^-,\sperp)
\end{align}
to write
\begin{align}
    &f^{+-} = -g f_{abc} t^c \intop_{\pperp, \qperp} \intop_{v^+}\intop_{v^-} \tilde \beta^{i,a}_A(x^+\! - v^+,\pperp) \nonumber \\
    &\times \tilde \beta^{i,b}_B(x^-\! - v^-, \qperp) J_0( |\pperp\! +\! \qperp | \tau')  e^{-\ii (\pperp+\qperp) \cdot \xperp}.
\end{align}
Transforming back to coordinate space
\begin{align}
        f^{+-} &= -g f_{abc} t^c \intop_{\pperp, \qperp} \intop_{v^+}\intop_{v^-} \intop_{\uperp, \sperp}   \beta^{i,a}_A(x^+\! - v^+, \uperp) \nonumber \\
        &\times \beta^{i,b}_B(x^-\! - v^-, \sperp)J_0( |\pperp\! +\! \qperp | \tau') e^{i (\pperp \cdot \uperp + \qperp \cdot \sperp)} e^{-\ii (\pperp+\qperp) \cdot \xperp}
\end{align}
we can write
\begin{align}
    \pperp \cdot \uperp + \qperp \cdot \sperp = \frac{1}{2}(\pperp + \qperp) \cdot (\uperp + \sperp) + \frac{1}{2} (\pperp - \qperp) \cdot (\uperp - \sperp).
\end{align}
Now we use $\kperp = \pperp + \qperp$ and $\Delta \kperp = \frac{1}{2}(\pperp - \qperp)$ to integrate out $\Delta \kperp$ (fixing $\sperp = \uperp$) and $\sperp$, which yields
\begin{align}
        f^{+-} &= -g f_{abc} t^c \intop_{\kperp} \intop_{v^+}\intop_{v^-} \intop_{\uperp}   \beta^{i,a}_A(x^+\! - v^+, \uperp)  \nonumber \\
        &\ind\times \beta^{i,b}_B(x^-\! - v^-, \uperp) J_0(|\kperp| \tau') e^{-\ii \kperp \cdot (\xperp - \uperp)}.
\end{align}

Next, we use $\kperp \cdot (\xperp - \uperp) = k |\xperp-\uperp| \cos \theta$ (where $k\coloneqq |\kperp|)$ and integrate out $\theta$, which gives another Bessel function
\begin{align}
        f^{+-} &= -gf_{abc} t^c \intop_{0}^{\infty} \frac{dk \, k}{2\pi} \intop_{v^+}\intop_{v^-} \intop_{\uperp}   \beta^{i,a}_A(x^+\! - v^+, \uperp) \nonumber \\
        &\ind\times \beta^{i,b}_B(x^-\! - v^-, \uperp) J_0(k \tau') J_0(k |\xperp-\uperp|).
\end{align}
Using the closure relation 
\begin{align} \label{eq:bessel_closure}
\intop_0^{\infty} dk\,kJ_\nu(ka)J_\nu(kb)=\frac{\delta(a-b)}{a}
\end{align}
we get
\begin{align}
        f^{+-} &= -\frac{g}{2\pi} f_{abc} t^c \intop_{v^+}\intop_{v^-} \intop_{\uperp}   \beta^{i,a}_A(x^+\! - v^+, \uperp) \nonumber \\
        &\ind\times \beta^{i,b}_B(x^-\! - v^-, \uperp) \frac{\delta(\tau' - |\xperp-\uperp|)}{\tau'}. 
    \label{eq:fpm_intermediate}
\end{align}
Note that
\begin{align}
    \intop_0^{\infty}d v^+\intop_0^{\infty}dv^- = \intop_{-\infty}^{\infty}d\eta'\intop_0^{\infty}d\tau'\,\tau'
\end{align}
with $\eta' = \ln(v^+/v^-)/2$ and $\tau' = \sqrt{2v^+v^-}$ as before. Performing this change of variables, the delta function in Eq.~\eqref{eq:fpm_intermediate} is removed by integrating over $\tau'$, which gives the result
\begin{align}
        f^{+-} &= -\frac{g}{2\pi} f_{abc} t^c \intop_{-\infty}^{\infty} d\eta' \intop_{\uperp}  \beta^{i,a}_A(x^+\! -\! \frac{|\xperp- \uperp|}{\sqrt{2}} e^{+\eta'}, \uperp) \nonumber \\
        &\ind \times \beta^{i,b}_B(x^-\! -\! \frac{|\xperp-\uperp|}{\sqrt{2}} e^{-\eta'}, \uperp).
\end{align}
Finally, we shift the integration variable $\uperp= \xperp - \vperp$, which yields
\begin{align}
        f^{+-}&=  -\frac{g}{2\pi} f_{abc} t^c \intop_{-\infty}^{\infty} d\eta' \intop_{\vperp}  \beta^{i,a}_A(x^+\! -\! \frac{|\vperp|}{\sqrt{2}} e^{+\eta'}, \xperp - \vperp)  \nonumber \\
        &\ind \times\beta^{i,b}_B(x^-\! -\! \frac{|\vperp|}{\sqrt{2}} e^{-\eta'}, \xperp - \vperp).
\end{align}

To deal with $f^{+i}$ we can use the $\tilde \beta^{i,a}_{A/B}$ defined before to write the terms in Eq.~\eqref{eq:fpi} as
\begin{align}
    \tilde \phi^a_A \tilde \phi^b_B p^i \qperp^2 &= \ii \tilde \beta^{i,a}_A \widetilde{ \p^j_{(s)}\beta^{j,b}_B}, \\
    \tilde \phi^a_A \tilde \phi^b_B (- q^i \pperp^2) &= - \ii \widetilde{\p^j_{(u)} \beta^{j,a}_A} \tilde \beta^{i,b}_B, \\    
    \tilde \phi^a_A \tilde \phi^b_B (-2  q^i \pperp \cdot \qperp) &= - 2 \ii \tilde \beta^{j,a}_A \widetilde{\p^i_{(s)} \beta^{j,b}_B},
\end{align}
where $\widetilde{(\dots)}$ refers to a Fourier transformation in the transverse plane. We can then write $f^{+i}$ as
\begin{align}
    f^{+i}&= g f_{abc} t^c \intop_{\pperp, \qperp, v} \tilde G^{i,ab}(x^+\!- v^+, x^-\!- v^-, \pperp, \qperp) \nonumber \\
    &\ind \times \frac{1}{|\pperp+\qperp|} \frac{v^+}{\tau'} J_1(|\pperp+\qperp| \tau') e^{-\ii(\pperp + \qperp)\cdot \xperp},
\end{align}
with
\begin{align}
    &\tilde G^{i,ab}(x^+\! - v^+, x^-\!- v^-, \pperp, \qperp) \equiv \nonumber \\
    &\qquad \tilde \beta^{i,a}_A \widetilde{\p^j_{(s)} \beta^{j,b}_B} - \widetilde{\p^j_{(u)}\beta^{j,a}_A }\tilde \beta^{i,b}_B - 2 \tilde \beta^{j,a}_A \widetilde{\p^i_{(s)} \beta^{j,b}_B}.
\end{align}
Next, we can Fourier transform the terms in $\tilde G^{i,ab}$ to get
\begin{align}
    f^{+i} &= g f_{abc} t^c \intop_{\pperp, \qperp, v} \intop_{\uperp, \sperp} G^{i,ab}(x^+\!- v^+, x^-\!- v^-, \uperp, \sperp) \nonumber \\
    &\ind\times \frac{1}{|\pperp+\qperp|} \frac{v^+}{\tau'} J_1(|\pperp+\qperp| \tau') e^{\ii(\pperp \cdot \uperp + \qperp \cdot \sperp)} e^{-\ii(\pperp + \qperp)\cdot \xperp}
\end{align}
with
\begin{align}
    G^{i,ab} &=  \beta^{i,a}_A  \p^j_{(s)}\beta^{j,b}_B -\p^j_{(u)} \beta^{j,a}_A  \beta^{i,b}_B - 2  \beta^{j,a}_A \p^i_{(s)} \beta^{j,b}_B.
\end{align}

Now we are in a position to repeat the same steps as before: introduce $\kperp = \pperp + \qperp$ and $\Delta \kperp = \frac{1}{2}(\pperp - \qperp)$, integrate out $\Delta \kperp$ (which sets $\sperp=\uperp$) and $\sperp$ and finally integrate out the angle $\theta$ between $\kperp$ and $\xperp - \uperp$ to get another $J_0$, which gives
\begin{align}
    f^{+i} &= \frac{g}{2\pi}  f_{abc} t^c \intop_0^{\infty} dk\intop_{v} \intop_{\uperp} G^{i,ab}(x^+\!- v^+, x^-\!- v^-, \uperp, \uperp) \nonumber \\
    &\ind \times \frac{v^+}{\tau'}J_0(k|\xperp-\uperp|) J_1(k \tau').
\end{align}
At this point, a key observation is that we can rewrite $G^{i,ab}$ as
\begin{align}
    G^{i,ab} &= \p^j_{(u)}(\beta^{i,a}_A \beta^{j,b}_B-\beta^{j,a}_A\beta^{i,b}_B -\delta^{ij}\beta^{k,a}_A\beta^{k,b}_B) \nonumber \\
    & \equiv \p^j_{(u)}G^{ij,ab},
\end{align}
where the derivative acts on the $\uperp$-dependence of $\beta_A$ and $\beta_B$ (which now depends on $\uperp$ instead of $\sperp$). Assuming that $G^{ij}\rightarrow 0$ as $|\uperp|\rightarrow \infty$ we can now partially integrate in $\uperp$ to obtain
\begin{align}
f^{+i} &= \frac{g}{2\pi}  f_{abc} t^c \intop_0^{\infty} dk\intop_{v} \intop_{\uperp} G^{ij,ab}(x^+\!- v^+, x^-\!- v^-, \uperp, \uperp) \nonumber \\
&\ind \times \frac{v^+}{\tau'}k\,w^jJ_1(k|\xperp-\uperp|) J_1(k \tau').
\end{align}
We have introduced the unit vector
\begin{align}
    w^j = \p^j_{(u)} |\xperp- \uperp| =  \frac{x^j- u^j}{|\xperp-\uperp|}.
\end{align}
The closure relation for the Bessel functions gives
\begin{align}
f^{+i} &= \frac{g}{2\pi}  f_{abc} t^c \intop_{v} \intop_{\uperp} G^{ij,ab}(x^+\!- v^+, x^-\!- v^-, \uperp, \uperp) \nonumber \\
&\ind \times \frac{v^+}{\tau'}w^j\frac{\delta(\tau'-|\xperp-\uperp|)}{\tau'}.
\end{align}
Integrating out the delta function and performing the shift $\uperp = \xperp - \vperp$ finally yields
\begin{align}
    &f^{+i}
    = \frac{g}{2\pi}  f_{abc} t^c\nn\\
    &\hspace{0.2cm}\times\intop_{\eta'} \intop_{\vperp} G^{ij,ab}(x^+\!\!-\! \frac{|\vperp|}{\sqrt{2}}e^{+\eta'}, x^-\!\!-\! \frac{|\vperp|}{\sqrt{2}}e^{-\eta'}, \xperp\!-\!\vperp, \xperp\!-\!\vperp) \nn\\
    &\hspace{1.5cm}\times\frac{e^{+\eta'}}{\sqrt{2}}w^j,
\end{align}
where the $\eta'$-integral goes from $-\infty$ to $\infty$. Note that $w^j=v^j/|\vperp|$. We now write
\begin{align}
    &f_{abc}t^cG^{ij,ab} = V^{ij}-\delta^{ij}V,
\end{align}
with $V$ and $V^{ij}$ already defined in Eqs.~\eqref{eq:def_V} and \eqref{eq:def_Vij}.
The $f^{-i}$ component can be worked out in the same way as $f^{+i}$, while $f^{ij}$ is analogous to $f^{+-}$. The results for all independent components of the perturbative field strength tensor are
the expressions given in Eqs.~\eqref{eq:f+-}--\eqref{eq:fij} in the main text.

\section{Longitudinal flow from superimposed Bjorken flow}\label{app:ueta}

\begin{figure}[t]
    \centering
    \includegraphics{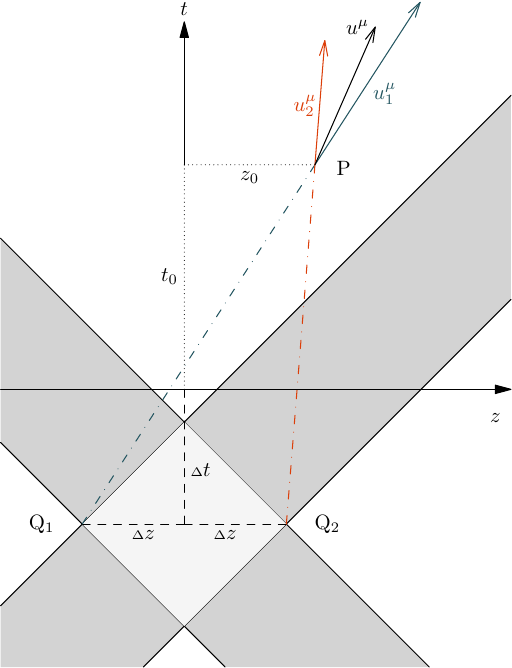}
    \caption{Geometric interpretation of the rapidity-dependence of the velocity. The Minkowski diagram shows the traces of the two colliding nuclei (gray) and their overlap, the collision region. At the $z$-symmetric points $\mathrm{Q}_1$ and $\mathrm{Q_2}$, Bjorken flow is produced and evolves to the point P\@. The two contributions are shown in terms of velocity vectors. The center arrow represents their energy-density weighted sum.}
    \label{fig:sub_bjorken}
\end{figure}

We construct a simple model to predict the $\eta_s$-dependence of $u^\eta$ from basic considerations. Our setup is shown in Fig.~\ref{fig:sub_bjorken}. Ignoring the transverse direction, we pick an arbitrary point $P$ in the forward light cone with Minkowski coordinates $(t_0, z_0)$. Note that the origin of the corresponding coordinate system is shifted with respect to the center of the collision region (the gray diagonal bars represent the traces of the nuclei). We start Bjorken flow at each point in the collision region and look at the resulting 4-velocity of the medium at $P$. We show two such contributions $u_1^\mu$ and $u_2^\mu$ coming from the points $Q_1 = (-\Delta t, -\Delta z)$ and $Q_2 = (-\Delta t, \Delta z)$ in the collision region. We weigh these contributions with the energy density $\epsilon_{1/2} = \epsilon_0/\tau_{1/2}$, with $\tau_{1/2}$ the proper time elapsed between $Q_{1/2}$ and $P$ and find
\begin{align}
\label{eq:eps_1_u_1}
    \epsilon_1 u_{1/2}^t &= \epsilon_0\frac{t_0 + \Delta t}{(t_0+\Delta t)^2 - (z_0\pm \Delta z)^2},\\
    \label{eq:eps_2_u_2}
    \epsilon_2 u_{1/2}^z &= \epsilon_0\frac{z_0 \pm \Delta z}{(t_0+\Delta t)^2 - (z_0\pm \Delta z)^2}.
\end{align}
We now take the sum $\epsilon u^\mu = \epsilon_1 u_1^\mu + \epsilon_2 u_2^\mu$ of these two contributions at $P$. We only consider the ratio $u^t/u^z$, for which we find
\begin{align}
    \frac{u^t}{u^z} = \frac{\frac{t_0 + \Delta t}{(t_0+\Delta t)^2 - (z_0+ \Delta z)^2}+\frac{t_0 + \Delta t}{(t_0+\Delta t)^2 - (z_0- \Delta z)^2}}{\frac{z_0 + \Delta z}{(t_0+\Delta t)^2 - (z_0+ \Delta z)^2}+\frac{z_0 - \Delta z}{(t_0+\Delta t)^2 - (z_0- \Delta z)^2}}> \frac{t_0}{z_0}
\end{align}
for positive $z$. We used $t_0>z_0$ and $\Delta t > \Delta z$ to evaluate this inequality. The case $z<0$ is analogous with the $>$ replaced by $<$. What we find is that the sum from the two ($z$-symmetric) Bjorken flow contributions, properly taking into account energy density, leads to a combined flow where $|u^t/u^z|$ is larger than it would be for Bjorken flow. This argument holds for any pair of points in the collision region and we thus conclude that the total flow at P must have negative $u^\eta$ at positive $z_0$. If P were located at negative $z_0$ or, equivalently, at negative $\eta_s$, then $u^\eta$ would be positive. Overall, the longitudinal expansion of the system is slower than for Bjorken flow.

We can formalize this argument further and consider the result of contributions from all over the collision region diamond, that is, we consider the integral
\begin{align}
\label{eq:eps_u_integral}
    \epsilon_\mathrm{tot} u^\mu_\mathrm{tot} = \intop_{\scalebox{1.2}{$\diamond$}} d\Delta t\, d\Delta z\, \epsilon(\Delta t, \Delta z)\, u ^\mu(\Delta t, \Delta z),
\end{align}
where the $\epsilon(\Delta t, \Delta z)u^\mu(\Delta t, \Delta z)$ are the analogues of Eqs.~\eqref{eq:eps_1_u_1} and \eqref{eq:eps_2_u_2} changing continuously over the collision region $\scalebox{1.5}{$\diamond$}$. These velocities, and consequently the result of the integral, depend on $t_0$ and $z_0$. After transforming to Milne coordinates and normalizing the total velocity, we obtain a result $u^\eta_\mathrm{tot}(\eta_s)$ for fixed $\tau$. We solve Eq.~\eqref{eq:eps_u_integral} analytically and show the resulting curve in Fig.~\ref{fig:ueta_curves}.

\bibliographystyle{elsarticle-num}
\bibliography{bib}

\end{document}